\documentclass[twocolumn]{aastex62}
\usepackage{multirow}
\usepackage{comment}
\usepackage{pifont}
\usepackage{amsmath}

\newcommand{\iras}{{\it IRAS}}
\newcommand{\wise}{{\it WISE}}
\newcommand{\spitzer}{{\it Spitzer}}
\newcommand{\herschel}{{\it Herschel}}
\newcommand{\kme}{{\tt KMESTM}}
\newcommand{\taumir}{$\tau_{9.7}$}

\newcommand{\ltorus}{$L_\mathrm{torus}$}

\newcommand{\lhost}{$L_\mathrm{galaxy}$}
\newcommand{\lco}{$L'_\mathrm{CO}$}
\newcommand{\lir}{$L_\mathrm{IR}$}
\newcommand{\umin}{$U_\mathrm{min}$}
\newcommand{\md}{$M_d$}
\newcommand{\mg}{$M_\mathrm{gas}$}
\newcommand{\ms}{$M_*$}
\newcommand{\qpah}{$q_\mathrm{PAH}$}
\newcommand{\hi}{H{\sevenrm\,I}}
\newcommand{\molh}{H$_2$}
\newcommand{\w}{$W$}
\newcommand{\mcl}{\multicolumn}
\newcommand{\ph}{\phantom}
\newcommand{\tnm}{\tablenotemark}
\newcommand{\ergps}{$\mathrm{erg\,s^{-1}}$}
\newcommand{\msunpy}{$M_\odot\:\mathrm{yr^{-1}}$}
\newcommand{\acounit}{$M_\odot\,\mathrm{(K\, km\, s^{-1}\, pc^{2})^{-1}}$}
\newcommand{\comh}{CO-to-$\mathrm{H_2}$}
\font\sevenrm=cmr7 scaled 1000

\submitjournal{ApJ}

\shorttitle{Interstellar Medium and Star Formation of LIRGs}
\shortauthors{Shangguan et al.}

\begin{document}

\title{Interstellar Medium and Star Formation of Starburst Galaxies on the 
Merger Sequence}

\correspondingauthor{Jinyi Shangguan}
\email{shangguan@pku.edu.cn}

\author[0000-0002-4569-9009]{Jinyi Shangguan}
\affil{Kavli Institute for Astronomy and Astrophysics, Peking University,
Beijing 100871, China}
\affiliation{Department of Astronomy, School of Physics, Peking University,
Beijing 100871, China}

\author[0000-0001-6947-5846]{Luis C. Ho}
\affil{Kavli Institute for Astronomy and Astrophysics, Peking University,
Beijing 100871, China}
\affiliation{Department of Astronomy, School of Physics, Peking University,
Beijing 100871, China}

\author[0000-0001-8496-4162]{Ruancun Li}
\affiliation{Department of Astronomy, School of Physics, Peking University,
Beijing 100871, China}

\author[0000-0001-5105-2837]{Ming-Yang Zhuang}
\affil{Kavli Institute for Astronomy and Astrophysics, Peking University,
Beijing 100871, China}
\affiliation{Department of Astronomy, School of Physics, Peking University,
Beijing 100871, China}

\author[0000-0002-9707-1037]{Yanxia Xie}
\affil{Kavli Institute for Astronomy and Astrophysics, Peking University,
Beijing 100871, China}

\author[0000-0001-5113-7558]{Zhihui Li}
\affil{Department of Astronomy, School of Physics, Peking University,
Beijing 100871, China}

\begin{abstract}
The interstellar medium is a key ingredient that governs star formation in 
galaxies.  We present a detailed study of the infrared ($\sim 1-500$ \micron) 
spectral energy distributions of a large sample of 193 nearby ($z \lesssim 
0.088$) luminous infrared galaxies (LIRGs) covering a wide range of 
evolutionary stages along the merger sequence.  The entire sample has been 
observed uniformly by 2MASS, \wise, \spitzer, and \herschel.  We perform 
multi-component decomposition of the spectra to derive physical parameters of 
the interstellar medium, including the intensity of the interstellar radiation 
field and the mass and luminosity of the dust.  We also constrain the presence
and strength of nuclear dust heated by active galactic nuclei.  The radiation 
field of LIRGs tends to have much higher intensity than in quiescent galaxies, 
and it increases 
toward advanced merger stages as a result of central concentration of the 
interstellar medium and star formation.  The total gas mass is derived from the 
dust mass and the galaxy stellar mass.  We find that the gas fraction of LIRGs
is on average $\sim$0.3 dex higher than that of main-sequence star-forming 
galaxies, rising moderately toward advanced merger stages.  All LIRGs have 
star formation rates that place them above the galaxy star formation main 
sequence. Consistent with recent observations and numerical simulations, the 
global star formation efficiency of the sample spans a wide range, filling the 
gap between normal star-forming galaxies and extreme starburst systems. 
\end{abstract}


\section{Introduction} 
\label{sec:intro}

Luminous infrared galaxies (LIRGs; Sanders \& Mirabel 1996), defined as systems
with total infrared (IR; 8--1000 \micron) luminosity \lir\ $>10^{11}\,L_\odot$,
\footnote{We use LIRGs to refer to all the galaxies with \lir\ $> 10^{11}\,
L_\odot$, although galaxies with \lir\ $> 10^{12}\,L_\odot$ are usually called 
ultraluminous infrared galaxies (ULIRGs).} have been studied extensively since they 
were recognized as a major constituent of the galaxy population from the
{\it Infrared Astronomical Satellite} (\iras) all-sky survey.  The power source
of the IR emission, whether it be star formation and/or active galactic nuclei 
(AGNs), has been intensively debated over the years (e.g., Genzel et al. 1998; 
Lutz et al. 1998; Spoon et al. 2007; Veilleux et al. 2009; Yuan et al. 2010; 
Iwasawa et al. 2011; Petric et al. 2011).  The star formation rate (SFR) of 
LIRGs, inferred from IR luminosity, generally exceeds $\gtrsim 10$ \msunpy, 
qualifying them as starburst systems that lie above the  SFR--\ms\ ``main 
sequence'' relation of low-$z$ star-forming galaxies (e.g., Daddi et al. 
2007; Noeske et al. 2007; Peng et al. 2010; Renzini \& Peng 2015).  Moreover, 
LIRGs dominate star formation at $z \gtrsim 1$ (e.g., Elbaz et al. 2002; 
Chapman et al. 2005).  Nearby LIRGs that are well-measured at a wide variety of 
wavelengths are, therefore, important to shed light on galaxy star formation 
at high redshifts.  Given their state of rapid stellar mass growth, LIRGs are 
important to study the coevolution of galaxies and their central supermassive 
black holes (Kormendy \& Ho 2013).  The most luminous members of the 
class---ULIRGs---may evolve into quasars (Sanders et al. 1988a, 1988b) and, 
finally, massive elliptical galaxies (Wright et al. 1990; Genzel et al. 2001; 
Tacconi et al. 2002) with the aid of AGN feedback (Di~Matteo et al. 2005; 
Hopkins et al.  2008; but see Shangguan et al. 2018, and references therein).  

The interstellar medium (ISM) is of great importance to understand the physics 
of star formation in LIRGs.  Early CO observations (e.g., Sanders \& Mirabel 
1985; Sanders et al. 1986, 1991; Young et al. 1986) revealed that LIRGs contain
large amounts of molecular gas, but that they emit IR emission in excess of 
the \lco--\lir\ relation of normal, star-forming galaxies.  Interferometric 
observations of some small samples of LIRGs with signatures of interactions 
found the CO emission mostly concentrated in the central regions (Downes \& 
Solomon 1998; Bryant \& Scoville 1999).  More recent, high-resolution 
observations show that the molecular gas in LIRGs is concentrated in compact 
central disks associated with nuclear starbursts (Ueda et al. 2014; Xu et al. 
2014, 2015; Scoville et al. 2015).  It has been widely debated whether 
starburst systems follow the same relation between SFR and gas content as 
regular, star-forming galaxies.  Kennicutt (1998a) argues that normal galaxies, 
starburst nuclei, and LIRGs obey the same empirical relation between SFR 
surface density and total gas (\hi+\molh) mass surface density.  However, more 
recent CO surveys suggest that normal galaxies and starburst systems behave 
differently in terms of their relation between the molecular gas and SFR 
(Daddi et al. 2010; Genzel et al. 2010), with the caveat that the conversion 
factor from CO emission to molecular gas mass (Bolatto et al. 2013) remains 
controversial (e.g., Liu et al. 2015).  

From a theoretical point of view, mergers and interactions are expected to 
efficiently drive gas inflow toward the galactic center, igniting a central 
starburst (e.g., Barnes \& Hernquist 1996; Mihos \& Hernquist 1996; Bournaud 
et al. 2011; Hopkins et al. 2013).  However, the gas kinematics in mergers are 
complex and comparison with model predictions is not straightforward (Iono et 
al. 2004a, 2005; Saito et al. 2015).  In the local Universe ($z \lesssim 0.1$),
the Great Observatories All-sky LIRG Survey (GOALS; Armus et al. 2009) provides
a complete sample of 201 LIRGs with observations from radio to X-rays.  This 
sample of IR-luminous galaxies span a diverse range of morphologies: 
non-mergers, pre-mergers, and mergers from early to late stage (Haan et al. 
2011; Petric et al. 2011; Stierwalt et al. 2013; Larson et al. 2016).  Objects
with the highest IR luminosities are primarily late-stage mergers (Sanders 
1988a; Dinh-V-Trung et al. 2001; Veilleux et al. 2002; Kim et al. 2013).  The 
diversity of merger stages encapsulated in GOALS is important to reveal the 
properties of the gas content along the evolutionary merger sequence.  
Yamashita et al. (2017) recently show that the size of the CO emission in the 
central kpc decreases from early- to late-stage mergers, while the molecular 
gas mass remains constant, statistically supporting the notion that gas inflow 
commonly replenishes nuclear starbursts in merging LIRGs.

We combine 2MASS, \wise, and \herschel\ photometric measurements to analyze the 
IR (1--500 \micron) spectral energy distributions (SEDs) of the entire GOALS 
sample.  We derive the mass and luminosity of the dust associated with the 
large-scale ISM of the host galaxy, after decomposing the emission from the 
hot dust powered by the AGN, and we place constraints on the intensity of the 
interstellar radiation field (ISRF).  The total gas mass and the SFR are 
then derived.  LIRGs tend to have moderately higher ISM fractions than normal, 
star-forming galaxies, possibly due to selection effects.  The large sample 
size and diverse morphologies of GOALS enables us to compare the distribution 
of physical properties as a function of different merger stages.  We find that 
the ISRF intensity, as probed by the galactic dust, increases toward the 
advanced merger stages, as does the ISM mass fraction.  LIRGs occupy a wide 
region above the main sequence of low-$z$ star-forming galaxies, 
with late-stage mergers exhibiting the highest SFRs.  The star formation 
efficiency (SFE), although spanning a wide range across the sample, tends to 
increase toward advanced merger stages.  LIRGs fills the bimodality of SFE 
previously found for normal and starburst systems.

This paper is organized as follows.  Section \ref{sec:data} describes the details 
of the sample and the data reduction.  We explain the methods to model 
the SEDs in Section \ref{sec:model} and present the SED fitting results in Section 
\ref{sec:sed}.  The stellar mass and ISM properties are presented in Section 
\ref{sec:results}.  Section \ref{sec:disc} discusses star formation in LIRGs.  We 
summarize the main conclusions in Section \ref{sec:conclude}.  This work adopts 
the following parameters for a $\Lambda$CDM cosmology: $\Omega_m = 0.308$, 
$\Omega_\Lambda = 0.692$, and $H_{0}=67.8$ km s$^{-1}$ Mpc$^{-1}$ (Planck 
Collaboration et al. 2016).

\section{Sample \& data reduction} 
\label{sec:data}

The 201 LIRGs from the GOALS sample are all mapped by the {\it Herschel Space 
Observatory} (Pilbratt et al. 2010) with both the Photodetector Array Camera and 
Spectrometer (PACS; Poglitsch et al. 2010), at 70, 100, and 160 \micron, and the 
Spectral and Photometric Imaging Receiver (SPIRE; Griffin et al. 2010), at 250, 
350, and 500 \micron.  Most of the merger systems are covered by single maps 
with each instrument.  Meanwhile, eight systems consist of widely separated 
pairs that require two PACS maps; their two components are measured separately.
Chu et al. (2017) provide integrated aperture photometry of the {\it Herschel}\
data for the entire GOALS sample.  We use their measurements of total 
integrated flux for the PACS and SPIRE data for all the 201 systems.  We 
supplement these data with our own measurements of near-IR photometry from
2MASS (Skrutskie et al. 2006) and mid-IR photometry from \wise\ (Wright et al. 
2010)\footnote{Only one of the two components in the eight widely separated 
systems is a LIRG.  We only consider the LIRG component and obtain the 
corresponding near-IR and mid-IR measurements.  The morphology of these objects
provided by Stierwalt et al. is also based on the LIRG component.} to construct
the full IR SED from 1 to 500 \micron.

We download the 2MASS and \wise\ images from the NASA/IPAC Infrared Science 
Archive (IRSA)\footnote{\url{irsa.ipac.caltech.edu/frontpage/}} and uniformly 
measure the integrated aperture photometry for the entire systems.  The details 
of our method, presented by Li et al. (2018, in preparation), are briefly 
summarized here.  A source mask is generated for each image based on the image 
segmentation file.  The sky background of the image is fitted with a
third-order polynomial function and subtracted; this suffices to remove 
the large-scale gradient in all of the 2MASS and \wise\ images.  We measure the
surface brightness profile of the targets by fitting isophotes with the 
IRAF\footnote{IRAF is distributed by the National Optical Astronomy 
Observatories, which are operated by the Association of Universities for 
Research in Astronomy, Inc., under cooperative agreement with the National 
Science Foundation.} task {\tt ellipse}.  The aperture in each band is determined 
separately by the isophote whose surface brightness reaches the large-scale 
variation of the background, which is estimated by sampling the rebinned 
background pixels.  One large aperture is used to enclose the entire 
merger system if the two galaxies are not coalesced, as we usually lack the 
resolution to separate the two galaxies in \herschel\ maps.  In order to provide 
aperture photometry consistent throughout the various IR bands, we choose 
the largest semi-major axis and semi-minor axis among all of the bands to arrive 
at the final aperture applicable to all the 2MASS and \wise\ images for each object.  
The final aperture sizes are always larger than the aperture adopted by Chu et al. 
for their \herschel\ measurements.  We omit five objects that are too large to be 
fully covered by 2MASS and three objects contaminated by bright stars, 
mostly in \w1 and \w2, resulting in the final sample of 193 LIRGs used in 
our current study.

Shangguan et al. (2018) showed that even SEDs that contain only photometric 
data from 2MASS, \wise, and \herschel\ can still yield robust cold dust masses 
and far-IR luminosities for the host galaxies of type 1 quasars.  However, the 
situation for some LIRGs is more complicated because of the strong effect of
silicate absorption features, which cannot be constrained well with photometric
data alone.  While \spitzer/IRS data exist for all GOALS objects, many of 
spectra cannot be directly used in the SED fitting because of their limited 
slit coverage.  Using a subset of 61 objects with IRS spectra that reasonably 
match the photometric data (Appendix \ref{apd:irs}), we show, by comparison 
of fits with and without inclusion of the spectra that the photometric SEDs 
alone can measure the interesting physical parameters without significant bias
(Appendix \ref{apd:spec}).

Table \ref{tab:sample} lists the basic information and the physical results 
from our photometric SED analysis of the sample of 193 LIRGs.  The luminosity 
distance is derived by correcting the heliocentric velocity from the galaxy 
peculiar motion using the 3-attractor flow model of Mould et al. (2000) and 
adopting our current cosmology for consistency.  We adopt the visually derived 
merger stage classification of Stierwalt et al. (2013), which is mainly based 
on {\it Spitzer}/IRAC 3.6 \micron\ images ($\sim$ 2\arcsec\ resolution) but 
complemented, whenever available, by high-resolution images from the 
{\it Hubble Space Telescope}\ (Haan et al.  2011).  They categorize the 
morphologies into five types: ``n'' for non-mergers (no signs of merger activity 
or massive neighbors), ``a'' for pre-mergers (galaxy pairs prior to a first encounter), 
``b'' for early-stage mergers (post-first-encounter with galaxy disks still symmetric 
but with signs of tidal tails), ``c'' for mid-stage mergers (showing amorphous 
disks, tidal tails, and other signs of merger activity), and ``d'' for late-stage 
mergers (two nuclei in a common envelope).  With additional ground-based 
optical images, Larson et al. (2016) compare their visual classifications to those 
of Stierwalt et al. (2013) for 65 objects in common and find reasonable 
consistency.  Due to limitations in resolution and sensitivity, stage ``b'' 
objects have a $\sim 50\%$ chance of being confused with stage ``a'' or ``c'';
stage ``d'' sources have $\lesssim 50\%$ chance of confusion with stage ``c'' 
and almost none with stage ``n''.  This level of uncertainty is, in fact, 
adequate for our purposes.

\section{SED models} 
\label{sec:model}

The SED fitting is conducted with a Bayesian Markov Chain Monte Carlo (MCMC) 
method developed by Shangguan et al. (2018).  The IR SED of a galaxy is 
dominated by stellar emission in the near-IR and dust emission at 
longer wavelengths.  We model the stellar emission as a 5 Gyr simple stellar 
population (Bruzual \& Charlot 2003; BC03), adopting a Chabrier (2003) initial 
mass function (IMF) and solar metallicity.  As the near-IR spectral shape of 
stellar emission is mostly governed by the old stellar population, it is 
relatively insensitive to stellar age.  Therefore, fixing the stellar age of 
the BC03 model will barely affect the SED fitting at longer wavelengths.  
Nuclear activity can produce prominent hot dust emission in the mid-IR, which 
can be modeled as a dusty torus.   We fit the cold dust emission from the host 
galaxy with the widely used physical dust model from Draine \& Li (2007; DL07),
which is based on the dust composition and size distribution observed in the 
Milky Way.  Two components of dust are considered: (1) most of the dust mass 
usually resides in the ``diffuse'' ISM exposed to the galactic ISRF with a 
minimum intensity $U=U_\mathrm{min}$; (2) a smaller mass fraction ($\gamma$) 
of the dust is associated with ``photo-dissociation regions'' heated by the 
ISRF with a power-law intensity distribution $U_\mathrm{min}<U<U_\mathrm{max}$.
The power-law index is fixed to $\alpha = 2$, and the maximum field intensity 
is set to $U_\mathrm{max} = 10^6$ (Draine et al. 2007).  The mass fraction of 
nanometer-size dust, a mixture of amorphous silicate and graphite, including 
polycyclic aromatic hydrocarbons (PAHs), is parameterized as $q_\mathrm{PAH}$.

For the objects that require a torus component, we adopt a new version of the 
CAT3D model (H{\"o}nig \& Kishimoto 2017) to fit the AGN dust torus emission.  
This model considers the different sublimation temperatures of silicate and 
graphite dust, self-consistently providing more emission from the hot dust at 
the inner edge of the torus, which was lacking in previous models such as 
CLUMPY (Nenkova et al. 2008a, 2008b), as well as in the earlier version of 
CAT3D (H{\"o}nig \& Kishimoto 2010).  Motivated by interferometric observations
(e.g., Raban et al. 2009), the new model can also include a wind component, 
which allows greater flexibility to accommodate the diversity of IR SEDs of 
quasars (Zhuang et al. 2018).  The basic CAT3D torus model consists five 
parameters: the inclination angle, $i$; the power-law index $a$ of the cloud 
radial distribution, of the form $r^a$, with $r$ the distance from the center 
in units of the sublimation radius $r_\mathrm{sub}$; the dimensionless scale 
height $h$ of the Gaussian distribution of clouds in the vertical direction, 
of the form $\exp\{-z^2/2(hr)^2\}$, with $z$ the vertical distance from the 
mid-plane; the average number $N_0$ of clouds along the equatorial 
line-of-sight; and the normalization factor $L$.  Limited by the degree of 
freedom, we use the basic CAT3D model to fit the photometric SEDs.  The fits 
that incorporate IRS spectra (Appendix \ref{apd:spec}) are conducted with an
additional wind component, which adds four additional free parameters: the 
radial distribution $a_w$ of dust clouds, the half-opening angle $\theta_w$, 
the angular width $\sigma_{\theta}$, and the wind-to-disk ratio $f_{\rm wd}$, 
which defines the ratio of the number of clouds along the cone and $N_0$.  
Garc{\'{\i}}a-Gonz{\'a}lez et al. (2017) also recently provide a new set of 
torus templates based on the CAT3D model.  We find that the choice of 
the torus model has little, if any, effect on measurements of cold dust 
properties.  Specifically, our various tests (see Appendix 
\ref{apd:spec}; Shangguan et al. 2018; Zhuang et al. 2018; Shangguan \& Ho
2018) find that the dust mass and \umin\ show scatter less than 0.1 and 0.15 
dex without significant systematic deviation.  We choose to use the results based on the 
templates from H{\"o}nig et al. (2017), as they provide the best overall 
fits (Zhuang et al. 2018). 

The silicate absorption at 9.7 \micron\ and 18 \micron\ in \spitzer/IRS spectra
indicates significant mid-IR extinction for a considerable fraction of our 
sample.  It is important to properly take into account dust extinction, as it 
affects not only the silicate features but also the overall shape of the 
broad-band continuum.  We adopt the dust extinction model of Smith et al. 
(2007).  The extinction model consists of a power law plus silicate features 
peaking at 9.7 and 18 $\mu$m, using the absorption properties of dust measured 
from the Milky Way.  Because the original extinction curve of Smith et al. 
(2007) ends at $\sim 38$ \micron, we extrapolate the curve to 1000 $\mu$m with 
a Drude profile with $\gamma_{r}=0.247$ peaking at 18 \micron, assuming no 
additional extinction features beyond 38 \micron\ (Mathis et al. 1990).  The 
only free parameter for the mid-IR extinction model is \taumir, the optical 
depth at 9.7 \micron.  

\section{SED fitting} 
\label{sec:sed}

With the models in hand, a key problem is whether the fits should include 
a torus component.  For most of the objects with relatively strong AGN-heated
dust emission, models without a torus component cannot fit the data.  However, 
most LIRGs have little if any obvious torus emission.  While many of the GOALS
objects have been previously studied in terms of their nuclear activity using 
a variety of multiwavelength diagnostics (e.g., Veilleux et al. 1999; Yuan et 
al. 2010; Iwasawa et al. 2011; Petric et al. 2011; Torres-Alb{\`a} et al. 2018), 
their AGN classification is not always clear because of complications from dust 
obscuration, strong star formation activity, and complex gas kinematics.  The 
mid-IR SED, on the other hand, is sensitive to the presence of the AGN dust 
torus (e.g., Stern et al. 2012; Blecha et al. 2018), such that highly 
obscured objects classified as non-AGNs by other methods may still show 
significant torus emission in the mid-IR (e.g., F00344$-$3349 and F01173+1405).  
In this study, we objectively ascertain whether a torus contribution is warranted 
based purely on the fitting results, not on any prior knowledge from other diagnostics.  
We fit the SEDs using models with and without a torus component and only 
choose the model with a torus component when the fit is significantly improved.  
Details of the fitting methods are reported in Section \ref{ssec:phot}.  In order 
to avoid model degeneracy, when the torus component is added, we fix 
$\gamma$ and \qpah\ of the DL07 component.  Thus, the fit is not always 
improved when the torus component is included.  In order to determine the 
best-fit model objectively, we calculate a local $\chi^2$ for the mid-IR region, 
using only the \w1 to \w4 bands.  Through various experimentation, we find 
that the torus component can be considered significant if including it in the 
fit reduces $\chi^2$ by more than a factor of 5.  We visually inspect every fit 
and in the end conclude that 69 (36\%) objects in the sample require 
a torus component.  Among these, it is noteworthy that 42 have been 
diagnosed previously as AGNs in the literature (Table 1), while two-thirds of 
the remaining 27 objects are likely AGNs according to their \wise\ color 
($W1-W2>0.5$; Mingo et al. 2016; Blecha et al. 2018).  We attempt to 
quantify the presence of a torus merely for the sake of completeness.  We 
emphasize that, as discussed in Shangguan et al. (2018), the properties of 
the cold dust derived from the DL07 component (e.g., $M_d$ and \umin) are 
actually very insensitive to whether or how the torus component is included 
in the fit.  Moreover, we find that the dust masses derived from full SED fitting 
are consistent with those obtained from fitting a modified blackbody (MBB) 
model to the FIR data only (Appendix \ref{apd:mbb}).

The mid-IR spectra of most of the sample only probe a fraction of the host 
galaxy due to the limited size of the IRS slit.  Using a subsample of 61 
objects whose IRS spectra are least affected by the problem of aperture 
mismatch, we show that the physical parameters of the DL07 model can be 
robustly derived from the photometric SED alone (Appendix \ref{apd:spec}).  
Most of the objects that show significant deviation between the photometric 
and full SED fitting can be identified from careful visual inspection of the 
photometric fits.  The unreliable fits usually show large, obvious mismatchs 
with the data or strong silicate absorption features, indicating that the 
model is poorly constrained by the data.

\subsection{Fitting the photometric SED} 
\label{ssec:phot}

We fit the SEDs using models with and without the torus component for all 193 
objects with robust near-IR to far-IR photometric measurements.  For fits 
without the torus component, we combine BC03 and DL07 components with the 
extinction applied to the latter.  The DL07 parameters \qpah, $\gamma$, \umin, 
and dust mass are all free.  When the torus component is included, we combine 
BC03, CAT3D, and DL07 components, again with extinction excluded from the 
stellar emission.  In view of the large number of free parameters and the very
limited number of \wise\ photometric data points, it is necessary to make 
some simplifying assumptions and keep certain parameters fixed.  We choose to 
use the CAT3D templates without the wind component, and for the DL07 component 
with $q_\mathrm{PAH}=0.47$ and $\gamma=0.03$, which are fiducial values 
found effective by Shangguan \& Ho (2018) for type 2 quasars.  
The model parameters are summarized in Table \ref{tab:pars}.

In the fitting process, the model SED is multiplied by the filter transmission
curve and integrated to calculate the average flux density of the various bands
(see Shangguan et al. 2018 for more details).  A considerable fraction of our
targets are (marginally) extended even in the \herschel/SPIRE bands.  The beam
size of the SPIRE bands varies with frequency, and the relative spectral
response function (RSRF)\footnote{According to Griffin et al. (2013), the 
transmission curve is the RSRF multiplied by the aperture efficiency.} 
effectively changes from point sources to extended sources.  Therefore, we 
need to evaluate the effect of the RSRF on our best-fit parameters of the DL07 
component, namely \umin\ and dust mass.  We select 30 objects that are mostly 
extended in SPIRE bands and fit their SEDs using the transmission curves for 
extended source.\footnote{The transmission curves are downloaded from the 
Spanish Virtual Observatory filter profile service: \url{http://svo2.cab.inta-csic.es/theory/fps3/}}  
Comparing to fits with the transmission curves of point sources, the dust mass 
and \umin\ are affected only at the level of $\sim$0.05 dex, with no obvious 
systematics.  Henceforth, we simply adopt the point-source RSRF.

\begin{figure*}
\begin{center}
\begin{tabular}{c c}
\includegraphics[height=0.2\textheight]{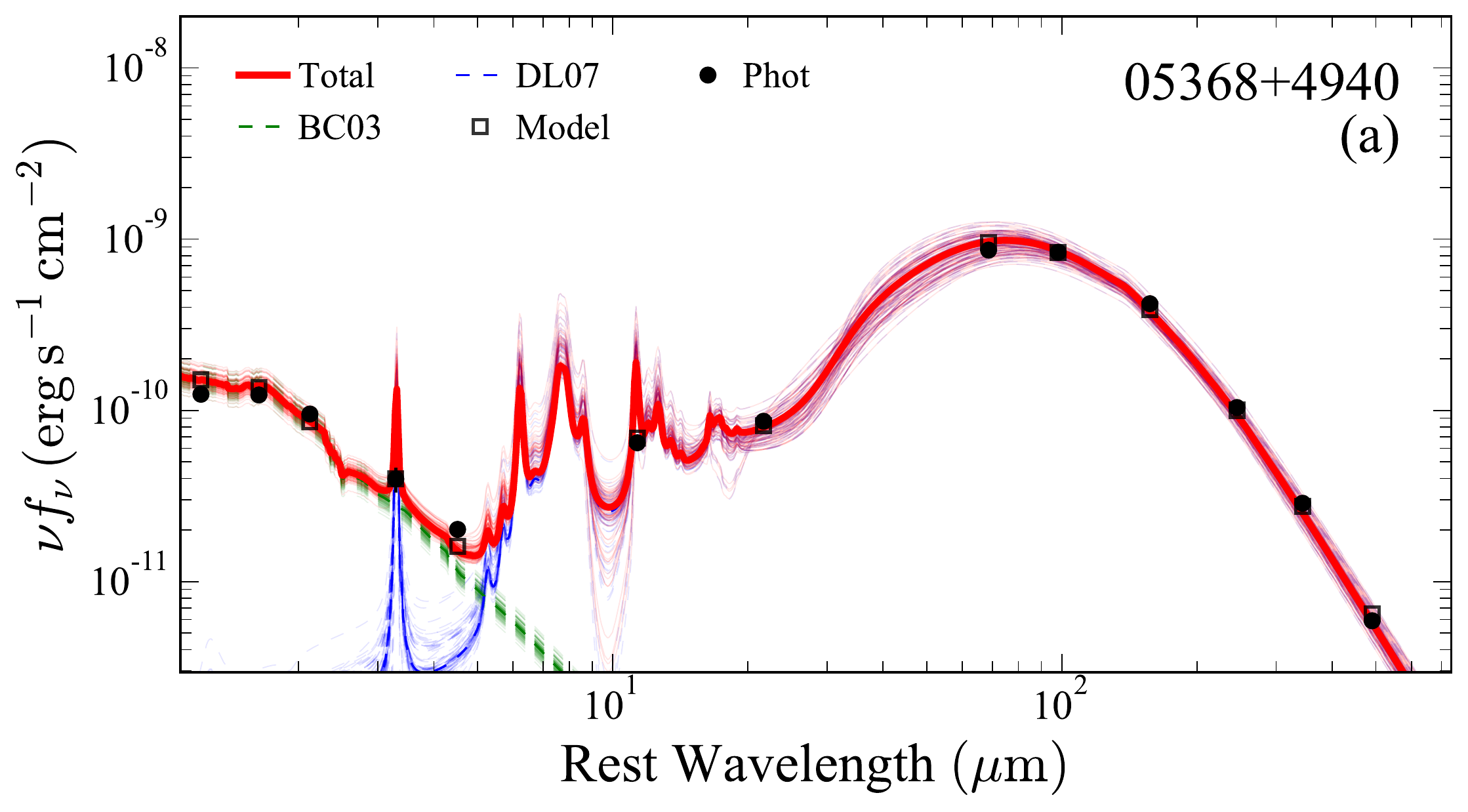} &
\includegraphics[height=0.2\textheight]{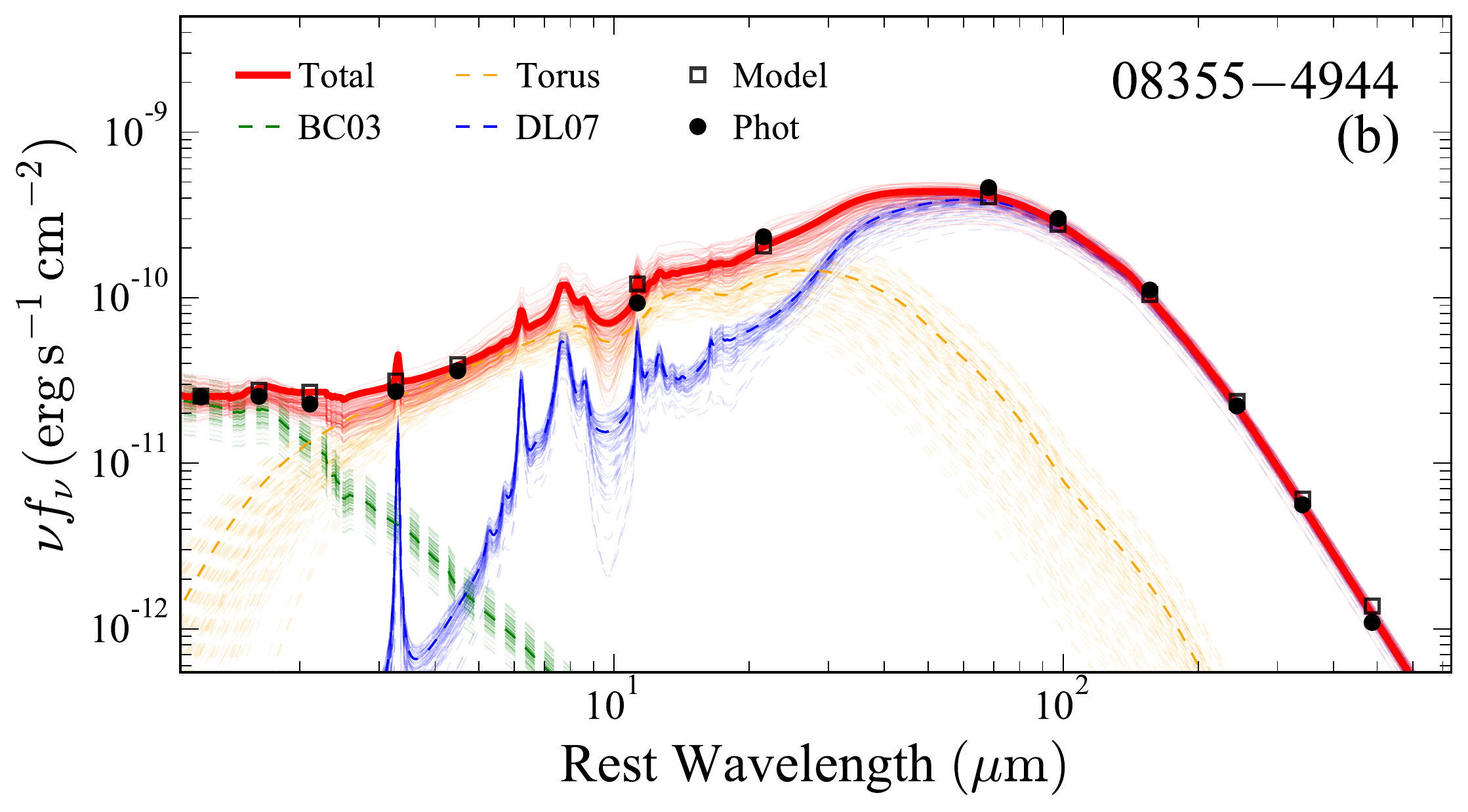} \\
\includegraphics[height=0.2\textheight]{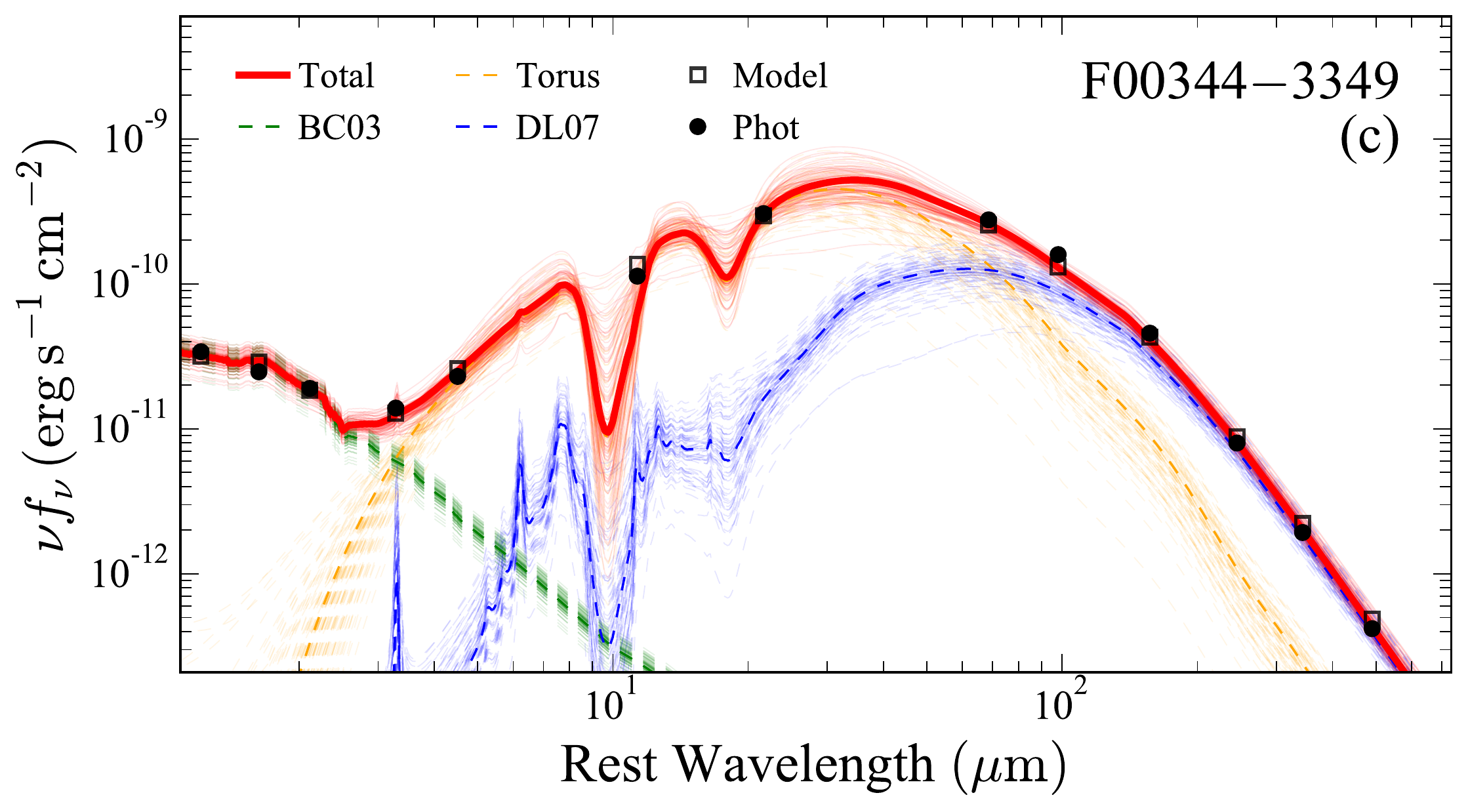} &
\includegraphics[height=0.2\textheight]{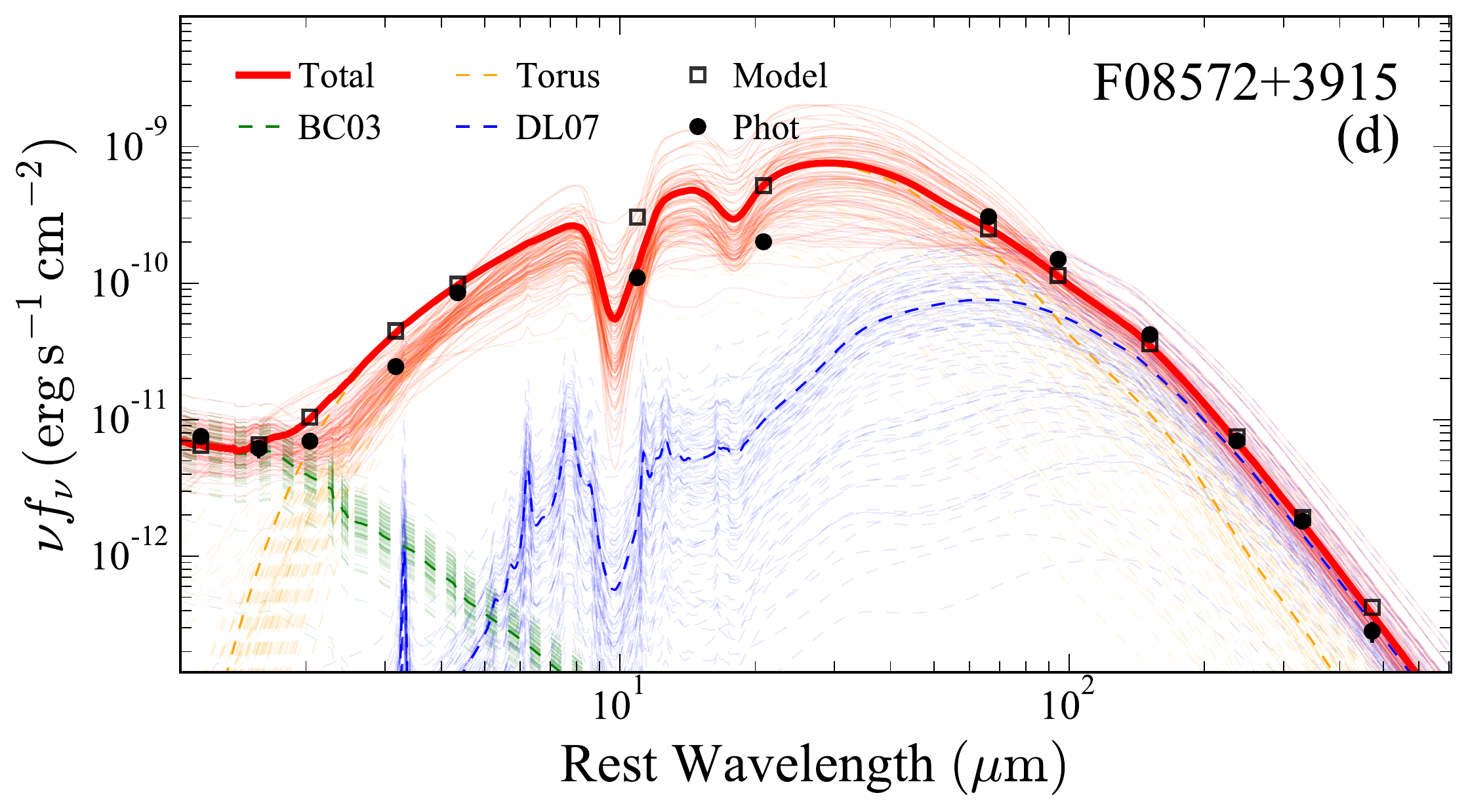} 
\end{tabular}
\caption{Examples of photometric SED fitting.  The black points 
are the photometric data from 2MASS, \wise, and \herschel.  The open squares 
are the model synthetic fluxes that can be directly compared with the data.  
The dashed lines are the individual components for stars (green; BC03), torus 
(orange; CAT3D), and host galaxy dust (blue; DL07).  The combined best-fit 
model is plotted as a red solid line.  To visualize the model uncertainties, 
the associated faint and thin lines represent 100 sets of models with 
parameters drawn randomly from the space sampled by the MCMC algorithm.  
}
\label{fig:phot}
\end{center}
\end{figure*}

Four examples of SED fits are shown in Figure \ref{fig:phot}, representing 
cases with low and high extinction, and low to moderate AGN torus contribution.
For cases like 05368+4940 (Figure \ref{fig:phot}a), which does not require a 
torus, the fit is very good.  In fact, the fits are generally robust even when 
the torus emission (Figure \ref{fig:phot}b) and/or the extinction (Figure
\ref{fig:phot}c) is significant.  However, the extinction cannot be 
accurately constrained when it is very strong, and the best-fit \umin\ and dust
mass may not be reliable.  Another problem is that the parameter range of 
\umin\ is limited to $\leq 25$, which is likely not high enough to fit some 
objects like F08572$+$3915.  As shown in Figure \ref{fig:spec}d, the best-fit 
model still lies below the 70 and 100 \micron\ data, such that in the 
photometric fit (Figure \ref{fig:phot}d) the torus component becomes very 
strong to compensate for the mismatch.  We visually check all the photometric 
SED fits and only find complications in 13 (7\%) of the cases; these 
are flagged in Table \ref{tab:sample}.  All objects whose photometric SED fits 
significantly deviated from the more robust fits using the IRS spectra 
(Appendix \ref{apd:spec}) are successfully identified by our visual 
inspection.

\section{Results}
\label{sec:results}

\subsection{Stellar mass} \label{ssec:stm}

The stellar mass is derived from the $J$-band photometry with a mass-to-light
ratio ($M/L$) constrained by the $B-I$ color (Bell \& de Jong 2001):

\begin{equation}\label{equ:sm}
\mathrm{log}\,(M_*/M_\odot) = -0.4 (M_J - M_{J, \odot}) - 0.75 + 0.34 (B - I),
\end{equation}

\noindent
where $M_J$ and $M_{J, \odot}=3.65$ (Blanton \& Roweis 2007) are the rest-frame 
$J$-band absolute magnitudes of the galaxy stellar emission and the Sun, 
respectively.  The IMF is converted from the ``scaled'' Salpeter (1955) value 
to that of Chabrier (2003) by subtracting 0.15 dex (Bell et al. 2003).\footnote{
Bell et al. (2003) provide the conversion from the ``scaled'' Salpeter (1995) 
IMF to the Kroupa et al. (1993) IMF, which is close enough to the Chabrier 
(2003) IMF (e.g., Madau \& Dickinson 2014).}  We calculate $M_J$ in two steps.
First, whenever the dust torus is included in the best-fit model, the torus 
contribution is removed from the $J$-band flux.  Then, K-correction is applied 
based on a 5 Gyr BC03 simple stellar population model assuming solar 
metallicity and a Chabrier (2003) IMF.  The uncertainty of the K-correction, 
considering the uncertainty of the star formation history, is $\sim 0.2$ mag.  
We adopt a constant color, $B - I = 2.0$ mag for LIRGs (Arribas et al. 2004; U 
et al. 2012).  The uncertainty of the color-based stellar mass is assumed 0.2 
dex (Conroy 2013).  The stellar masses of of the GOALS LIRGs range from $M_* = 
10^{10.1}\,M_\odot$ to $10^{11.5}\,M_\odot$, with a median value of 
$10^{10.9 \pm 0.3}\,M_\odot$.  All of the merger stages have a similar 
distribution of $M_*$.  As discussed in Appendix \ref{apd:mstar}, our stellar 
masses are broadly consistent with those given by Howell et al. (2010), given 
the relatively large uncertainty of $M/L$.

\subsection{Interstellar radiation field} 
\label{ssec:isrf}

\begin{figure}
\begin{center}
\includegraphics[height=0.35\textheight]{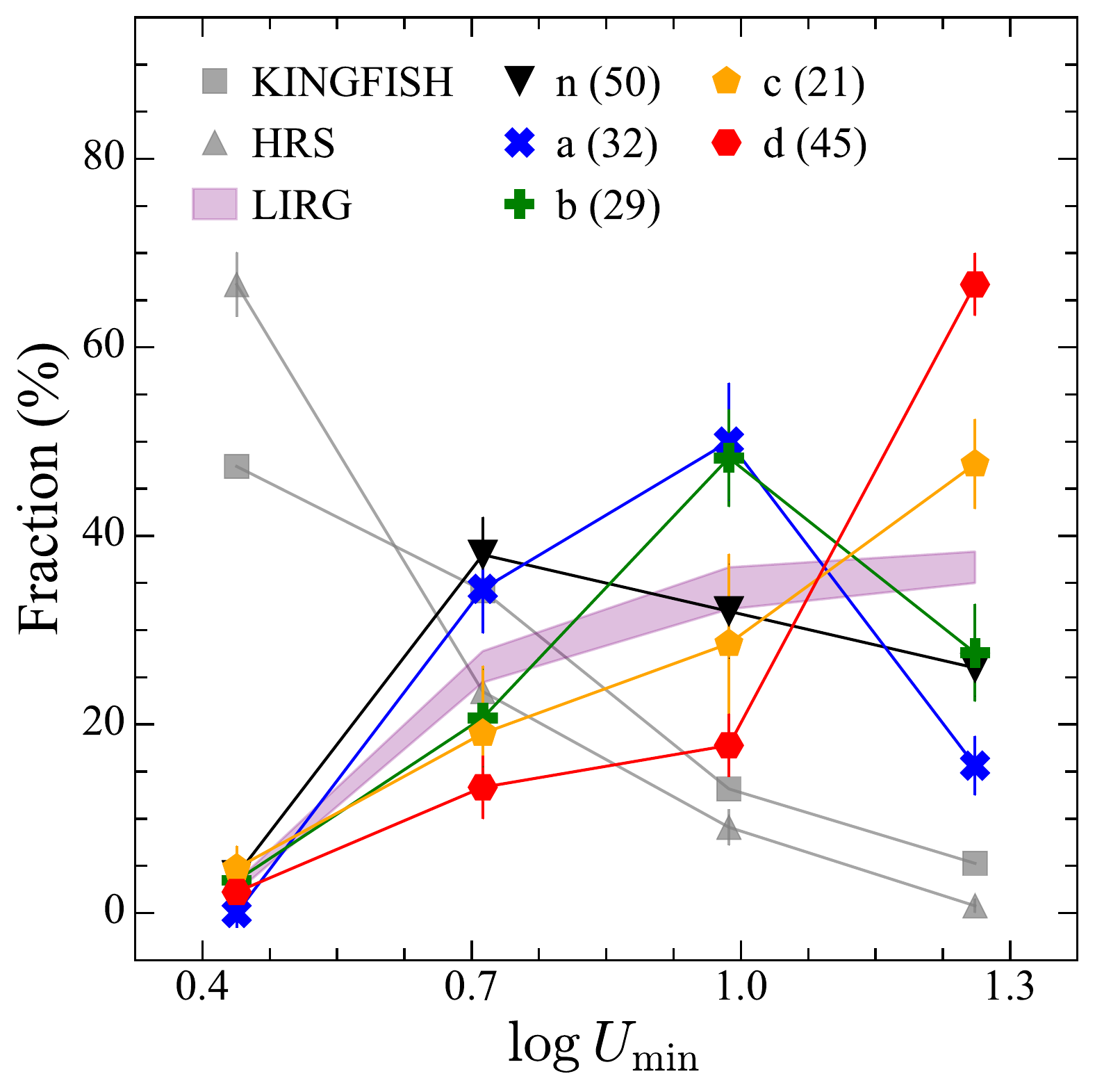}
\caption{The distributions of \umin\ in LIRGs are compared with star-forming 
galaxies from HRS (upward triangles; Boselli et al. 2010; Ciesla et al. 2014) 
and KINGFISH (squares; Kennicutt et al. 2011; Draine et al. 2007).  The the 
entire LIRG sample is shown in the purple shaded ($\pm 1\sigma$) region.  
The star-forming and quenched galaxies in the HRS and KINGFISH samples 
decrease toward high \umin, while the LIRGs tend to peak at high values.  
The  peak \umin\ of the LIRGs increases from non-mergers (``n'', black) to 
late-stage mergers (``d'', red).  The total number of objects at each merger 
stage with robust SED fits are listed in the legend.  The uncertainties are estimated 
with a Monte Carlo method, resampling the parameters according to their 
measurement uncertainties and calculating the number of galaxies in each bin 
500 times.  No error bars are associated with the KINGFISH galaxies because 
uncertainties are not available for them.
}
\label{fig:umin}
\end{center}
\end{figure}

The parameter \umin, mainly determined by the peak of the far-IR SED, probes 
the minimum intensity of the ISRF.  As all of our targets are well detected in 
the far-IR, our SEDs should be able to constrain \umin\ robustly, except 
perhaps for some objects with \umin\ $>25$ limited by the available parameter 
space of the DL07 templates.  LIRGs tend to have higher values of \umin\ than 
normal, star-forming galaxies (Figure \ref{fig:umin}).  Moreover, it is clear 
that \umin\ is generally higher toward late-stage mergers.  The elevated values
of \umin\ in LIRGs is likely due to their highly concentrated star formation 
and centrally peaked ISM distribution (da Cunha et al. 2010).  In support of 
this interpretation, submillimeter observations show high gas surface densities
within the central $\sim 1$ kpc of IR-luminous galaxies (Iono et al. 2004b; 
Ueda et al. 2014; Xu et al. 2014, 2015).  Although AGNs can heat the dust even
on global, galactic scales (e.g., Symeonidis 2017; Shangguan et al. 2018), the 
far-IR luminosity in most LIRGs is not likely dominated by AGNs (Genzel et al.
1998).  Within the GOALS sample, less than 50\% of the objects in each merger 
stage are diagnosed with AGN activity on the basis of our SED fitting or 
other diagnostics.

\subsection{ISM mass} 
\label{ssec:mass}

The dust masses range from $M_d = 10^{7.0}\,M_\odot$ to $10^{8.8}\,M_\odot$,
with a median value of $10^{8.2 \pm 0.3}\,M_\odot$.  We estimate the gas mass 
following $\log\,M_\mathrm{gas} = \log\,M_d + \log\,\delta_\mathrm{GDR} + 0.23$,
where the gas-to-dust ratio $\delta_\mathrm{GDR}$ is estimated from the galaxy 
stellar mass (Shangguan et al. 2018).\footnote{As discussed in Shangguan et al. 
(2018), a correction of 0.23 dex is applied to account for the extended \hi\ 
gas in the outskirts of the galaxy.}  The corresponding gas masses therefore span 
$M_\mathrm{gas} \approx 10^{9.3} - 10^{10.9}\,M_\odot$, with a median value of 
$10^{10.3 \pm 0.3}\,M_\odot$.  F03164$+$4119, a radio-loud AGN, is the only 
object with $\log\,(M_\mathrm{gas}/M_*)<-1.5$ or $\log\,(M_d/M_*)<-3.5$, 
significantly lower than the rest of the sample.  

It is not trivial to verify the reliability of our dust-based gas 
masses, as direct \hi\ measurements are lacking for most of our sample.
Nevertheless, we tried to compare our results with molecular gas masses for the 
subsample of 46 GOALS objects with CO measurements compiled by Larson et al. 
(2016), with the major caveat that the
molecular-to-total gas fraction is unknown.  Our total gas 
masses\footnote{We do not apply the 0.23 dex correction here, since it mainly 
accounts for \hi\ gas on the outskirts of the galaxy.} are consistent with 
the molecular gas masses, with a median difference of $0.09\pm0.24$ dex.  
Taken at face value, this might indicate that the molecular gas is able to 
account for most of the gas in the region of dust emission.  However, 
we have to emphasize that the \comh\ conversion factor adopted in 
Larson et al. (2016) is $X_\mathrm{CO} = 3.0 \times 10^{20}\, 
\mathrm{H_2\, cm^{-2}\, (K\, km\, s^{-1})^{-1}}$ or $\sim 6.5$ \acounit, which 
is $\sim 1.5$ times of the fiducial Milky Way value of $\sim 4.3$ \acounit\ 
(e.g., Bolatto et al. 2013) and $\sim 8$ times the value found in ULIRGs, 
$\sim 0.8$ \acounit.  Therefore, this test still suffers considerable uncertainty 
due to the \comh\ conversion factor.

\begin{figure*}
\begin{center}
\includegraphics[height=0.35\textheight]{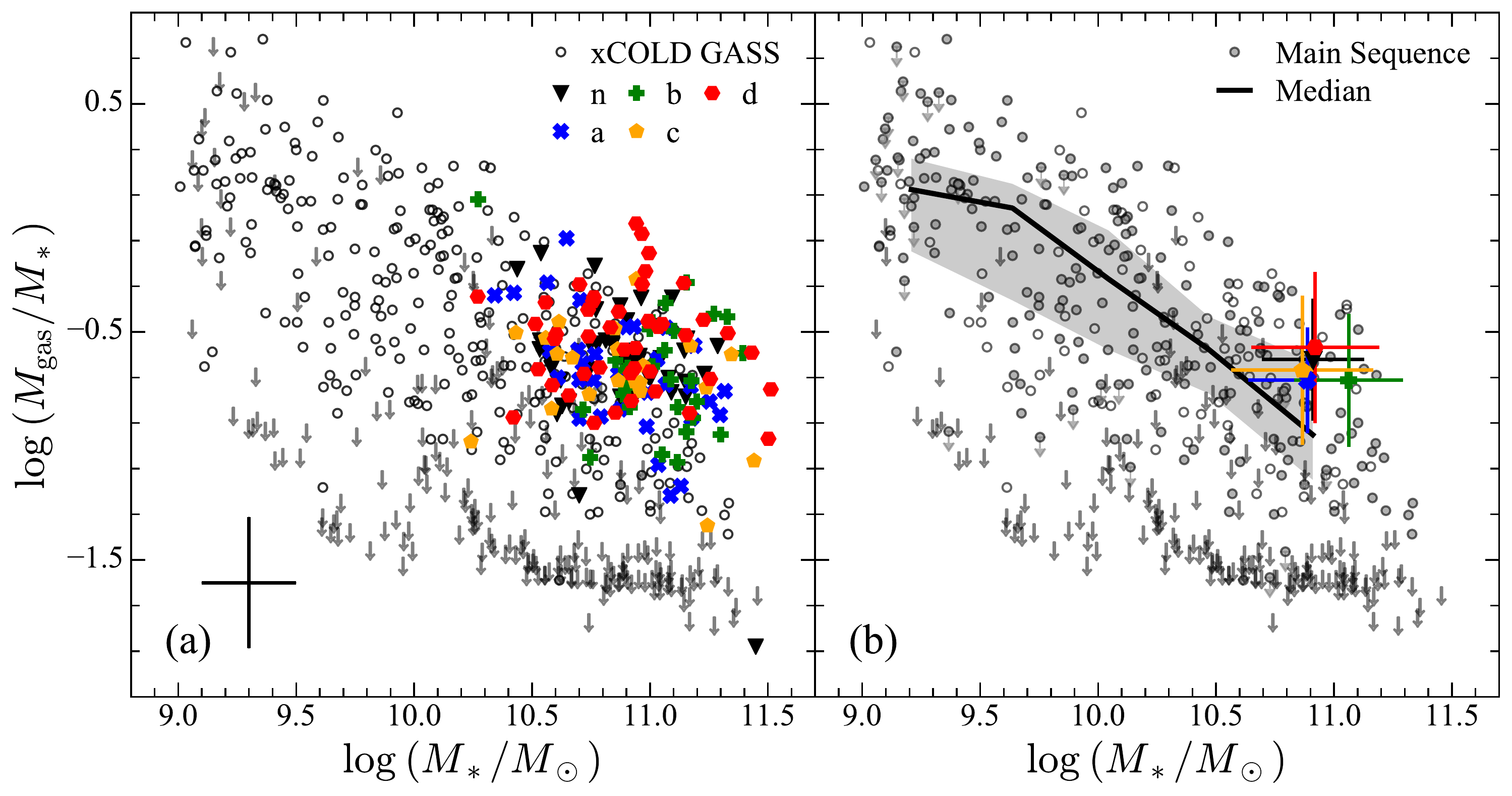}
\caption{The gas mass fraction of LIRGs in different merger stages are 
compared with the xCOLD GASS galaxy sample.  Individual objects are plotted in 
(a), while the $50^{+25}_{-25}$th percentile of the gas fractions at different 
merger stages are plotted in (b).  LIRGs in different merger stages are denoted 
following Figure \ref{fig:umin}.  The galaxies with measured total gas are 
denoted with empty circles, and those with upper limits are denoted with 
downward arrows. Galaxies on the main sequence are denoted with filled gray 
circles in (b).  The main sequence galaxies (Figure \ref{fig:ms}) are selected 
within $\pm 0.4$ dex of the relation provided by Saintonge et al. (2016; their 
Equation 5).  The median (thick line) and 25th-to-75th percentile (gray shaded 
region) of gas fraction of the main-sequence galaxies in (b) are calculated 
with \kme\ including the upper limits.  LIRGs tend to have moderately higher 
gas fraction than main-sequence galaxies.
}
\label{fig:mass}
\end{center}
\end{figure*}

Figure \ref{fig:mass}a compares the gas mass fraction of LIRGs with normal 
galaxies from xCOLD GASS.\footnote{xCOLD GASS (Saintonge et al. 2017) 
is a representative, mass-selected ($M_* > 10^9\,M_\odot$) sample of 532 
local ($0.01 < z < 0.05$) galaxies with both CO($1-0$) and \hi\ measurements.} 
LIRGs at different merger stages largely occupy a similar region in parameter 
space, with later merger stages preferentially exhibiting somewhat higher gas 
mass fractions.
By contrast, LIRGs as a group tend to have higher gas fractions than the 
overall xCOLD GASS sample.  This is not unexpected, for LIRGs are mostly 
starburst systems.  Typical main-sequence galaxies (Section \ref{ssec:sfr}) 
offer a more appropriate comparison.  In Figure \ref{fig:mass}b, we calculate 
the $50^{+25}_{-25}$th percentiles of the gas fraction of main-sequence 
galaxies, taking into account upper limits using the Kaplan-Meier product-limit
estimator \kme\ from ASURV (Feigelson \& Nelson 1985; Lavalley et al. 1992).  
LIRGs have moderately higher ($\sim 0.3$ dex) higher gas fractions than the 
median gas fraction of main-sequence galaxies in the xCOLD GASS sample.  

\begin{figure*}
\begin{center}
\includegraphics[height=0.35\textheight]{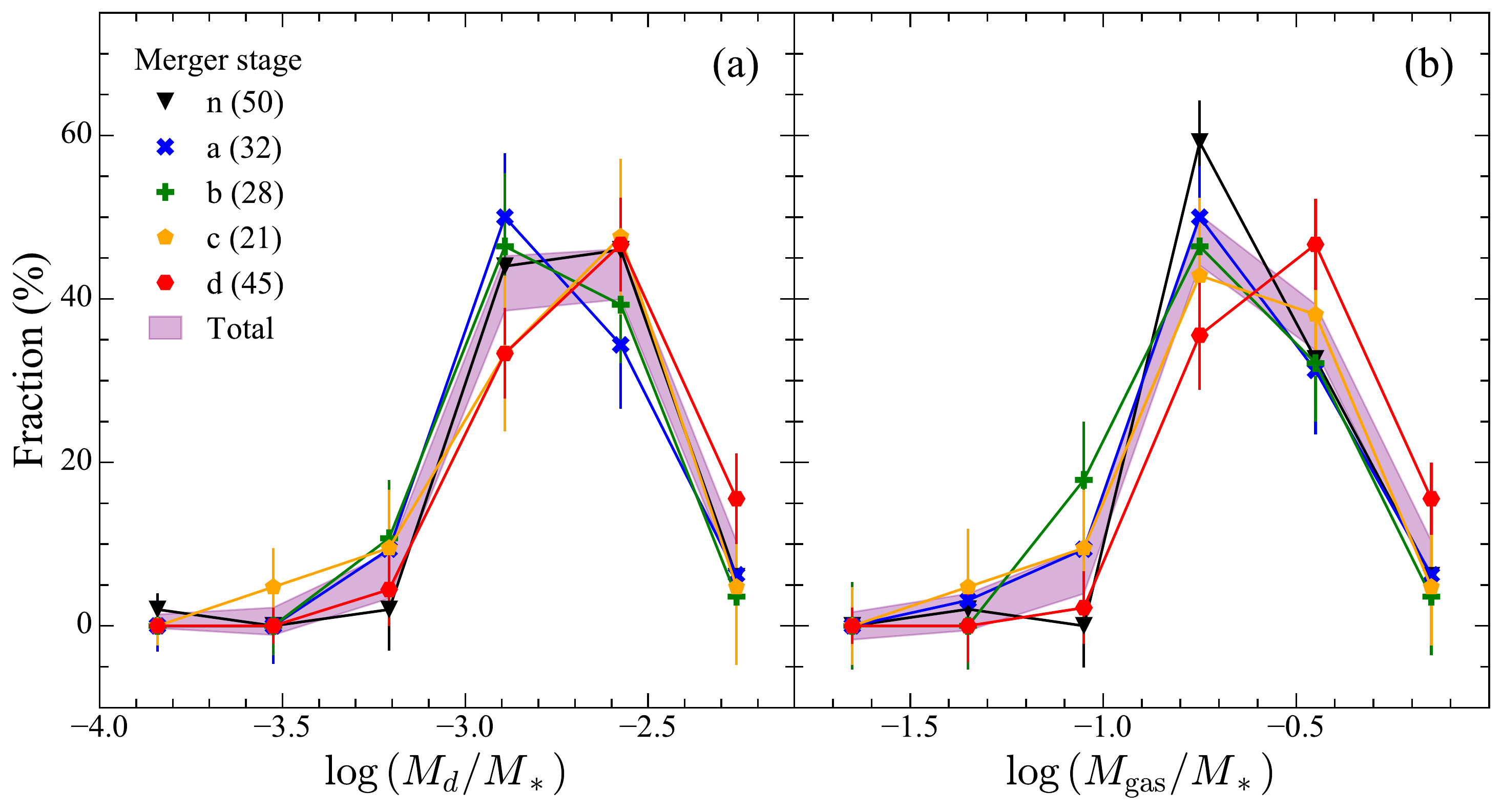}
\caption{The distribution of (a) dust and (b) gas mass fractions at different 
merger stages.  There is a trend for the gas fraction to increase toward 
late-merger stages.  The gas fraction of the late-stage mergers tends to be 
significantly different from the other subgroups.  The symbols and 
uncertainties are the same as in Figure \ref{fig:umin}.
}
\label{fig:fgas}
\end{center}
\end{figure*}

Divided into different phases along the merger sequence (Figure \ref{fig:fgas}),
the gas mass fraction of LIRGs tends to rise from the pre-merger stage (``a'') 
to the late-merger stage (``d'').  According to the Kolmogorov-Smirnov 
statistic, the ``d'' sample differs statistically significantly from samples ``a'' and 
``c'' ($p<0.05$), but, formally, not from sample ``b'' ($p<0.1$).  As discussed 
in Appendix \ref{apd:mbb}, the increase of gas fraction toward late-stage 
mergers holds also for dust masses derived from the modified blackbody (MBB) 
analysis.  

What is the physical origin of the gas enhancement?  Since our gas masses 
are inferred indirectly from dust emission, perhaps the apparent rise in gas 
mass fraction is an artifact of enhanced dust production in the nuclear 
starbursts of late-stage mergers (Haan et al. 2013).  However, whether 
starbursts lead to the preferential production or destruction of dust grains 
is unclear (Gall et al. 2011a, 2011b).  The effect, in any case, is only 
mild, as the gas fractions of LIRGs are only moderately higher than those of 
main-sequence galaxies.  Galaxy-galaxy mergers may increase the supply of cold 
gas through cooling of hot halo gas (Moster et al. 2011; Hwang \& Park 2015; 
Karman et al. 2015), but the observational evidence of enhanced gas mass 
fractions in galaxy pairs and post-merger galaxies is not clear-cut
(Ellison et al. 2015, 2018; Violino et al. 2018)

We end this section with a caveat.  Recall that our gas mass 
estimates depend critically on $\delta_\mathrm{GDR}$, which ultimately 
is tied to the mass--metallicity relation of isolated, star-forming galaxies 
(Tremonti et al. 2004; Kewley et al. 2008).  However, galaxy mergers in 
general (e.g., Ellison et al. 2008; Scudder et al. 2012) and LIRGs in 
particular (e.g., Rupke et al. 2008; Kilerci Eser et al. 2014) lie systematically 
below the mass-metallicity relation, by $\sim 0.2$ dex  (Herrera-Camus 
et al. 2018).  The metallicity of gravitationally disturbed systems are likely 
diluted by the inflow of more pristine, low-metallicity gas (Torrey et al. 2012; 
Bustamante et al. 2018).  Taken at face value, a reduction of 0.2 dex in 
metallicity in LIRGs will lead to an increase of 
$\delta_\mathrm{GDR}$ by the same factor because the two quantities 
are correlated almost linearly (Leroy et al. 2011; Magdis et al. 2012).  On 
the other hand, the formalism of Shangguan et al. (2018) to convert dust 
mass into total gas mass explicitly corrects $\delta_\mathrm{GDR}$ by a 
factor of 0.23 dex to account for \hi\ gas from the outskirts of the galaxy.  
This factor, in fact, is almost exactly identical to the metallicity offset for 
LIRGs reported by Herrera-Camus et al. (2018), strongly corroborating 
the scenario that dynamical interactions drive metal-poor \hi\ gas from 
the outskirts to the center of the galaxy.  Our methodology, in other 
words, is fully applicable to LIRGs despite possible variations in the 
mass--metallicity relation in these systems.  Note, further, the flatness 
of the mass--metallicity relation at the massive end implies that 
increasing the stellar mass by a factor of 2 only increases the metallicity 
by $< 0.1$ dex, much smaller than other uncertainties.

\section{Discussion}
\label{sec:disc}

\subsection{Star formation rate and the effect of AGNs} 
\label{ssec:sfr}

The cold dust emission we measure from fitting the photometric SED with the 
DL07 model presumably emanates from the large-scale ISM of the galaxy.  We 
denote as \lhost\ the $8-1000\,\micron$ integrated luminosity of this component.
Although extinction is formally taken into consideration in our fits, it has an 
almost negligible effect on the DL07 component.  The LIRGs with robust fits 
have \lhost\ $\approx 10^{44.2}$ to $10^{46.0}$ \ergps, with a median value of
$10^{45.0 \pm 0.3}$ \ergps.  Following Kennicutt (1998b; Equation 4), these 
values of \lhost\ translate to SFRs = 5.9$-$296 \msunpy\ (median 27 \msunpy); 
in accordance with other conventions throughout this paper, the SFRs refer to 
a Chabrier (2003) IMF.

\begin{figure*}
\begin{center}
\includegraphics[height=0.35\textheight]{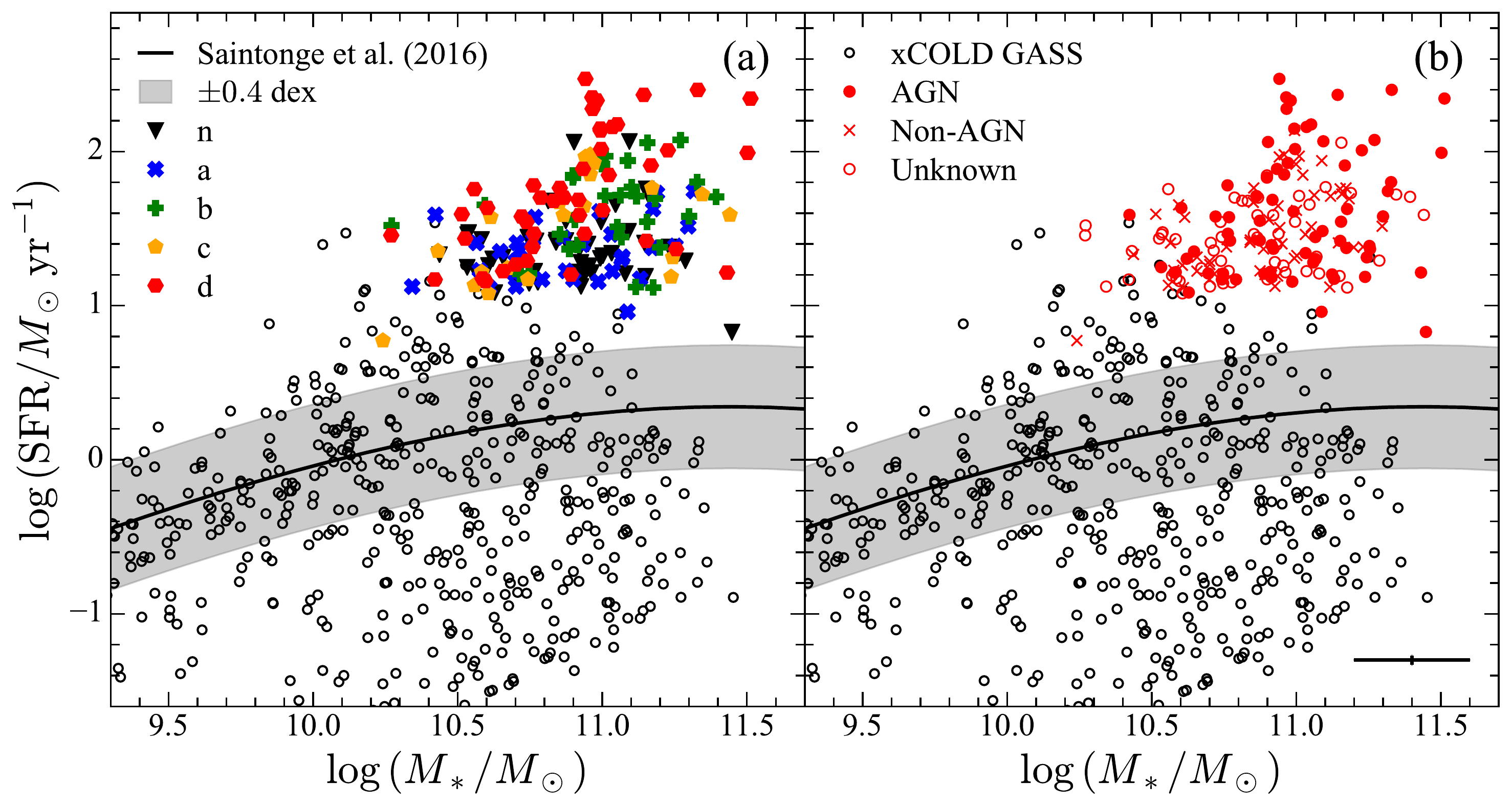}
\caption{LIRGs lie above the galaxy main sequence (solid line: Saintonge et 
al. 2016).  LIRGs in different merger stages are denoted following Figure 
\ref{fig:umin}, and galaxies from the xCOLD GASS sample are plotted as empty 
circles.  The shaded region is $\pm 0.4$ dex above and below the solid line, 
indicating the range of the main sequence.  In (b), LIRGs are divided into 
AGNs, non-AGNs, and unknown based on all the available diagnostics in the 
literature (Table \ref{tab:sample}).  The median error bars are shown in 
the lower-right corner.
}
\label{fig:ms}
\end{center}
\end{figure*}

As shown in Figure \ref{fig:ms}a, our sample of LIRGs are located 
systematically above the galaxy main-sequence, which, for consistency, is 
represented by the parametric relation of Saintonge et al. (2016) and by 
galaxies whose SFRs lie within $\pm 0.4$ dex (Chang et al. 2015) of the 
relation.  We note that Saintonge et al. (2016) define the main sequence based 
on SFRs derived from UV and mid-IR (12 or 22 \micron) luminosities of Sloan 
Digital Sky Survey DR7 galaxies with $0.01 < z < 0.05$ and $M_*>10^8\,M_\odot$.
It is still debatable whether the main sequence flattens beyond $M_* \approx 
10^{10.5}\,M_\odot$.  The detailed form of the galaxy main sequence depends on 
the selection criteria for star-forming galaxies (Renzini \& Peng 2015), as 
well as on the methodology used to derive SFRs (e.g., H$\alpha$ luminosity:
Peng et al. 2010; Renzini \& Peng 2015; UV+IR luminosity: Whitaker et al. 
2012; Lee et al. 2015; Saintonge et al. 2016).

Except for those with the highest SFRs ($\gtrsim100\,M_\odot\,\rm{yr^{-1}}$) 
and highest stellar masses ($M_* \gtrsim 10^{11}\,M_\odot$), 
which are almost exclusively late-stage mergers, LIRGs of different merger 
stages largely overlap with each other.  Considerable uncertainty surrounds 
the SFRs in LIRGs, however.  AGN contamination of FIR-based SFRs remains a 
possibility (Shangguan et al. 2018).  Moreover, AGNs may be hidden by strong 
dust obscuration, especially in late-stage mergers (e.g., Arp 220; Scoville et 
al. 2017).  Objects with identifiable AGN signatures do not stand out clearly 
from those that do not in Figure \ref{fig:ms}b, except that, as with the 
merger stage, nearly all sources with SFRs $\gtrsim100\,M_\odot\,\rm{yr^{-1}}$
and $M_* \gtrsim 10^{11}\,M_\odot$ are identified as AGNs.

\subsection{Star formation efficiency} 
\label{ssec:sfe}

The gas content of star-forming galaxies correlates strongly with their SFR 
(the Kennicutt-Schmidt relation; Kennicutt 1998a, and references therein).
It has been suggested that normal and starburst galaxies follow two different 
sequences of the Kennicutt-Schmidt relation, both for gas traced through 
lines (Daddi et al. 2010; Genzel et al. 2010) and indirectly through dust 
emission (Rowlands et al. 2014).  Rowlands et al.'s study combines local 
($z < 0.5$) dusty galaxies from the Herschel-Astrophysical TeraHertz Large 
Area Survey (H-ATLAS) and $z \approx 2$ submillimeter galaxies.  An 
important consequence of this result is that for a given amount of gas, 
starbursts generate stars with greater star formation efficiency, 
$\mathrm{SFE} \equiv \mathrm{SFR} / M_\mathrm{gas}$.  Not all 
investigators accept the reality of this apparent bimodality in SFE, as the 
result depends on the uncertainty of the \comh\ conversion factor (e.g., 
Narayanan et al. 2012), the exact formulation of the star formation law 
(Krumholz et al. 2012), and possible selection effects impacting the CO 
observations (Sargent et al. 2014).  

\begin{figure*}
\begin{center}
\includegraphics[height=0.35\textheight]{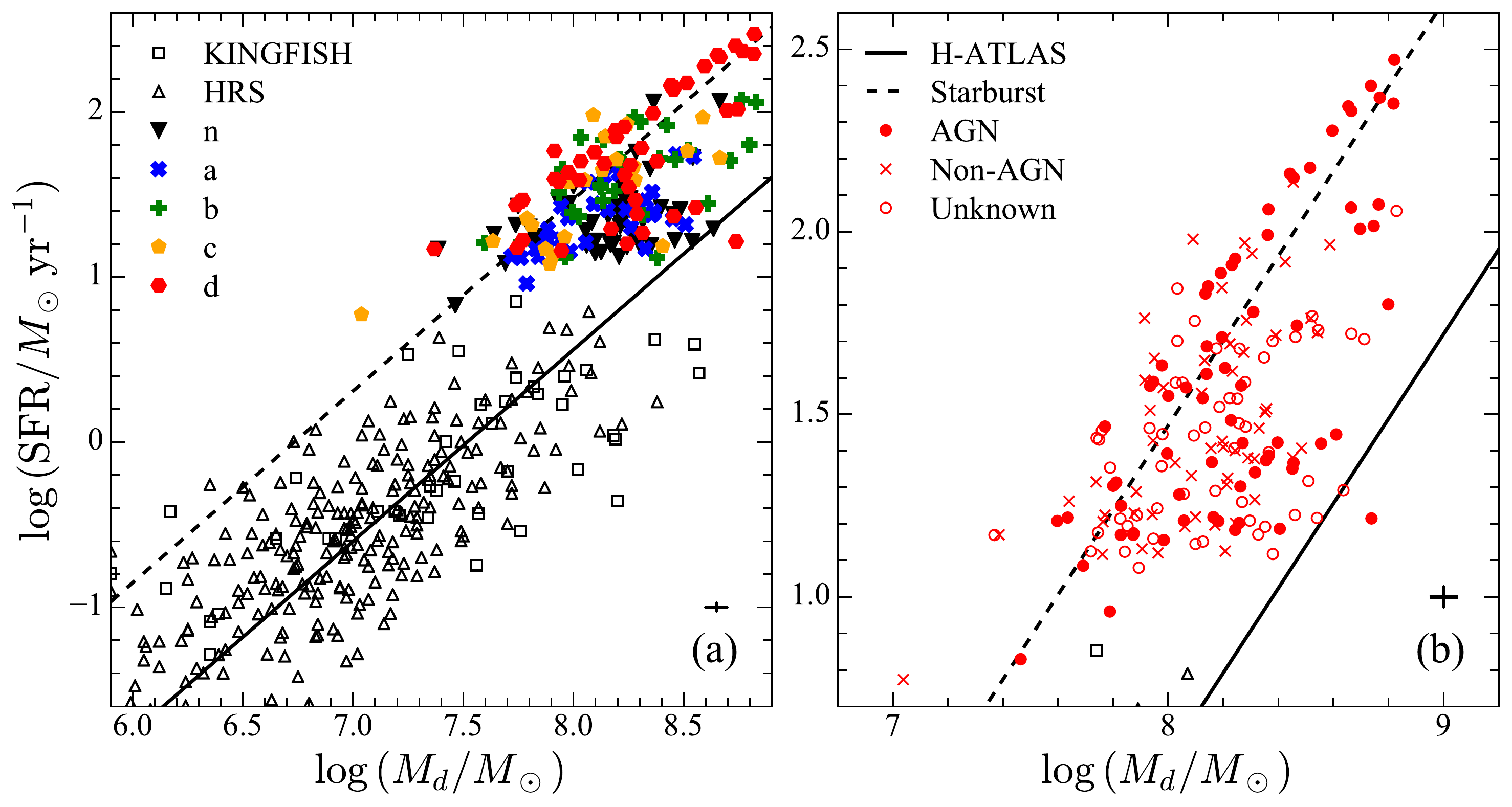}
\caption{LIRGs tend to have higher SFRs for the same dust mass compared to 
normal galaxies from the HRS and KINGFISH samples.  LIRGs in 
different merger stages are denoted following Figure \ref{fig:umin}, while 
normal galaxies are empty circles.  Panel (b) is zoomed in to better show the 
distribution of the AGNs, non-AGNs, and unknown subsamples of the LIRGs (Table 
\ref{tab:sample}).  The solid and dashed lines are the relations derived from 
low-$z$ dusty galaxies (H-ATLAS) and starburst galaxies (local ULIRGs and $z 
\approx 2$  submillimeter galaxies) by Rowlands et al. (2014), respectively.  
The median uncertainties are shown on the lower-right corner of each panel.
}
\label{fig:sfe}
\end{center}
\end{figure*}

Our new analysis of the GOALS sample provides a fresh opportunity to 
re-examine this issue, using our homogeneously derived, robust estimates 
of the SFRs and ISM masses.  Figure \ref{fig:sfe}a shows that the GOALS 
LIRGs lie essentially {\it in between}\ the two sequences of normal and starburst 
galaxies defined by Rowlands et al. (2014), in excellent agreement with the 
behavior of starbursts, as suggested recently (e.g., Saintonge et al. 2011; 
Sargent et al. 2014; Violino et al. 2018).
Different merger stages cannot be clearly distinguished, except for the handful
of the most extreme advanced mergers with the highest SFRs 
($\gtrsim100\,M_\odot\mathrm{yr^{-1}}$), which also possess the highest SFEs.  
We zoom in get a better view in Figure \ref{fig:sfe}b, now further highlighting
the AGNs.  Again, apart from the subset of the most dust-rich systems with the 
most extreme levels of star formation activity, AGNs do not stand out notably.
It is worth noting that the correlation between dust mass and SFR does 
not arise trivially from their mutual dependence on the IR emission.  This issue has been tested
by Santini et al. (2014) using mock SEDs of galaxies that cover the 
parameter space of dust mass and SFR of our LIRGs.  This is mainly because 
the far-IR emission is much more sensitive to the dust temperature than to the 
dust mass.

\begin{figure}
\begin{center}
\includegraphics[height=0.35\textheight]{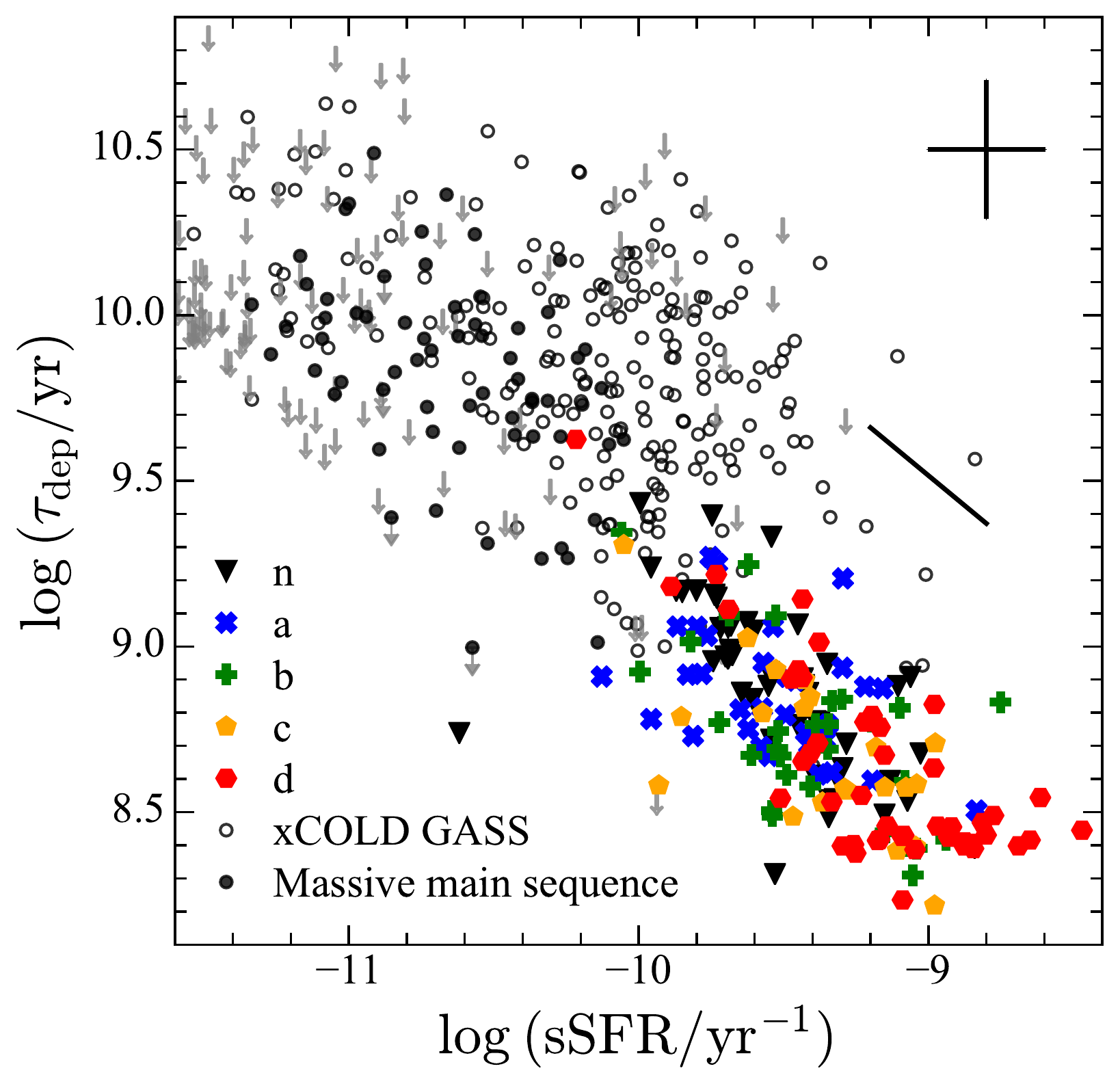}
\caption{The depletion timescale of the total gas and the specific SFR (sSFR) 
of LIRGs are compared with galaxies from xCOLD GASS sample.  LIRGs in different
merger stages are denoted following Figure \ref{fig:umin}.  The empty circles 
and downward arrows are galaxies from xCOLD GASS with detections and upper 
limits of the total gas mass, respectively.  The filled circles highlight the 
main-sequence galaxies with stellar mass $>10^{10.5}\,M_\odot$.  For 
comparison, the solid black line indicates the slope for the dependence of 
molecular gas depletion timescale on sSFR from Saintonge et al. (2011): 
$\tau_\mathrm{dep}(\mathrm{H_2}) \propto \mathrm{sSFR}^{-0.724}$.
The median uncertainties are shown on the upper-right corner.  LIRGs and 
massive main-sequence galaxies tend to follow a trend with a slope close to 
that of the molecular gas depletion timescale.
}
\label{fig:tdep}
\end{center}
\end{figure}

Lastly, Figure \ref{fig:tdep} illustrates the dependence of the gas depletion 
timescale, $\tau_{\rm dep} \equiv {\rm SFE}^{-1}$, with the specific SFR, 
${\rm sSFR} = {\rm SFR}/M_*$.  LIRGs usually have $\tau_{\rm dep} < 3$ Gyr, 
systematically shorter than most normal galaxies.  LIRGs and main-sequence
galaxies of similar stellar mass ($>10^{10.5}\,M_\odot$; black points) clearly 
follow a trend that is close to the relation between molecular gas depletion 
timescale and sSFR derived by Saintonge et al. (2011): 
$\tau_\mathrm{dep}(\mathrm{H_2}) \propto \mathrm{sSFR}^{-0.724}$.  LIRGs in
different merger stages largely overlap with each other along the trend, 
indicating that the SFR is not enhanced exclusively during any particular 
merger stage, although the more advanced stages (``c'' and ``d'') do tend to 
have systematically shorter $\tau_\mathrm{dep}$ and higher sSFR.  But, there 
are exceptions.  Advanced mergers with long $\tau_\mathrm{dep}$ and low 
sSFR do exist.  A few late-stage mergers with low SFE have markedly low dust 
temperatures (e.g., $T_d = 23.4 \pm 0.5$ K in F02070+3857; $T_d = 20.8 \pm 0.4$ 
K in F05365+6921).
It is conceivable that much of the ISM in these systems, despite being advanced 
mergers, has not yet settled to the center of the galaxy to fuel a nuclear 
starburst.  Depending on the details of the progenitor galaxies and the 
particulars of the orbital parameters, galaxy-galaxy interactions can enhance 
gas density and produce extended, clumpy star formation (Powell et al. 2013; 
Renaud et al. 2014) prior to the onset of a nuclear starburst, which is only 
triggered after sufficient inflow of cold gas occurs (e.g., Di~Matteo et al. 
2007; Hopkins et al. 2013).  In the opposite extreme, we also find non-mergers 
with high SFE (e.g., F23135$+$2517, F06592$-$6313, F14179$+$4927); 
these are perhaps triggered by minor rather than major mergers.

\section{Conclusions} 
\label{sec:conclude}

The sample of LIRGs in GOALS encompasses a large, homogeneously selected 
sample of nearby ($z \lesssim 0.088$), luminous (median $L_{\rm IR} = 
10^{11.4\pm0.3}\,L_\odot$), massive (median $M_*=10^{10.9\pm0.3}\,M_\odot$) 
star-forming galaxies covering a diverse range of morphologies representing 
different stages of the galaxy merger sequence.  We use a recently developed 
Bayesian MCMC fitting method to derive global physical parameters (total dust 
mass, gas mass, extinction, stellar mass, SFR, SFE) for 193 of the 201 GOALS 
galaxies by modeling their integrated IR ($1-500$ \micron) SEDs constructed from 
2MASS, \wise, and \herschel\ photometry.  The spectral decomposition of the SEDs 
also yields useful constraints on the intensity of the ISRF as well as the presence 
and strength of an AGN torus.  Using a subsample of objects with 
aperture-matched \spitzer/IRS spectra, we demonstrate that robust physical 
parameters can be measured with the wavelength coverage and quality of the 
available photometric data.  We use these measurements to investigate the 
ISM content and SFRs of LIRGs, with emphasis on their evolutionary phase 
along the merger sequence and the effect of AGNs.

The main conclusions are as follows:
 
\begin{enumerate}

\item As expected from their IR selection, LIRGs are rich in dust (and hence 
gas).  The dust masses of LIRGs range from $10^{7.0}$ to $10^{8.8}\,M_\odot$,
with a median of $10^{8.2 \pm 0.3}\,M_\odot$; these correspond to total (atomic
plus molecular) gas masses of $10^{9.3}$ to $10^{10.9}\,M_\odot$ (median 
$10^{10.3 \pm 0.3}\,M_\odot$).  The gas mass fractions (\mg/\ms) of LIRGs are
$\sim 0.3$ dex higher than those of normal, star-forming galaxies on the 
main sequence, the most gas-rich being those morphologically classified as 
late-stage mergers.
 
\item LIRGs have systematically stronger ISRFs than normal, star-forming 
galaxies.  Moreover, the intensity of the ISRF increases gradually but 
progressively from early- to late-stage mergers, likely a reflection of 
elevated ISM concentration and central star formation in advanced stages of 
the merger sequence.
 
\item The integrated IR ($8-1000\,\micron$) luminosities traced by the global, 
cold dust component implies SFRs of 5.9 to 296 \msunpy\ (median 27 \msunpy), 
placing these LIRGs systematically above the galaxy star-forming main sequence.
While different merger stages and levels of AGN activity show no strong 
correlation with location on the main sequence, objects with $\mathrm{SFR} 
\gtrsim 100$ \msunpy\ and $M_* \gtrsim 10^{11}\,M_\odot$ are all advanced, 
late-stage mergers with unambiguous signatures of AGNs.
 
\item LIRGs fill the gap in the bimodal distribution of SFEs previously defined 
by normal star-forming galaxies and starburst galaxies.  Advanced mergers
tend to exhibit systematically higher SFEs, while the variation of SFE is 
large among all the merger stages.  LIRGs obey and extend toward higher masses 
the trend between molecular gas depletion timescale and specific SFR defined 
by main-sequence galaxies.

\end{enumerate}

\acknowledgments

We are grateful to an anonymous referee for helpful comments and suggestions. 
J.S. thanks Jing Wang for helpful discussions.  
This research was supported by the National Key Program for Science and 
Technology Research and Development (2016YFA0400702) and the National Science 
Foundation of China (11303008, 11473002).  Y.X. is supported by China 
Postdoctoral Science Foundation Grant 2016 M591007.  This publication makes use 
of data products from the Two Micron All-Sky Survey, which is a joint project 
of the University of Massachusetts and the Infrared Processing and Analysis 
Center/California Institute of Technology, funded by the National Aeronautics 
and Space Administration and the National Science Foundation.  It also makes 
use of Astropy, a community-developed core Python package for astronomy
(Astropy Collaboration et al. 2013).

%

\vspace{5mm}


\software{Astropy (Astropy Collaboration et al. 2013);
          SciPy (Jones et al. 2001).
          }



\clearpage
\appendix

\section{{\it Spitzer}/IRS spectra} 
\label{apd:irs}

The mid-IR IRS spectra provide crucial information to constrain the 
torus and cold dust emission.  Although IRS spectra exist for the entire GOALS
sample, a major difficulty is that most of our targets are relatively nearby 
galaxies that have spatial extents larger than the IRS slit widths (3\farcs7 
for SL and 10\farcs6 for LL).  We visually checked the slit coverage of each 
object using available optical and near-IR images and conclude that 61 of the 
GOALS galaxies do not suffer from this aperture mismatch problem.
We obtain their IRS low-resolution spectra from the NASA/IPAC Infrared Science 
Archive.\footnote{\url{https://irsa.ipac.caltech.edu/data/GOALS/galaxies.html}}
The spectra were extracted with the standard extraction aperture and point 
source calibration modes in SPICE (Stierwalt et al. 2013).  The different 
orders of the SL and LL spectra are matched.  Since the slit width of SL is 
much smaller than that of LL, the flux level of the SL spectra is usually 
lower than that of the LL spectra, and the former needs to be scaled up to 
match the latter based on their overlapping region (Stierwalt et al. 2013).  
However, this method does not always provide an optimal scaling factor because 
the overlapping region may be affected by emission/absorption features and 
sometimes by the poor signal-to-noise ratio of the edge of the spectra.  
Therefore, we fine tune the scaling factor of SL by eye to achieve $\sim 5$\% 
accuracy.  We further scale the internally adjusted IRS spectra to match the 
integrated flux density of the \wise\  \w4 band.  The scaling factors of 
both steps are listed in Table \ref{tab:irs}; they are usually less than 30\%.  
Rigorously speaking, the aperture mismatches among SL, LL, and the integrated 
photometry may introduce systematic uncertainties into the SED fitting.  
However, this problem is beyond the scope of this work. Appendix \ref{apd:spec} 
tests the consistency between the SED fits with and without IRS spectra and
demonstrates that the currently available data are sufficiently accurate for 
our main goals.

\section{SED Fits with IRS Spectra} 
\label{apd:spec}

\begin{figure*}
\begin{center}
\begin{tabular}{c c}
\includegraphics[height=0.2\textheight]{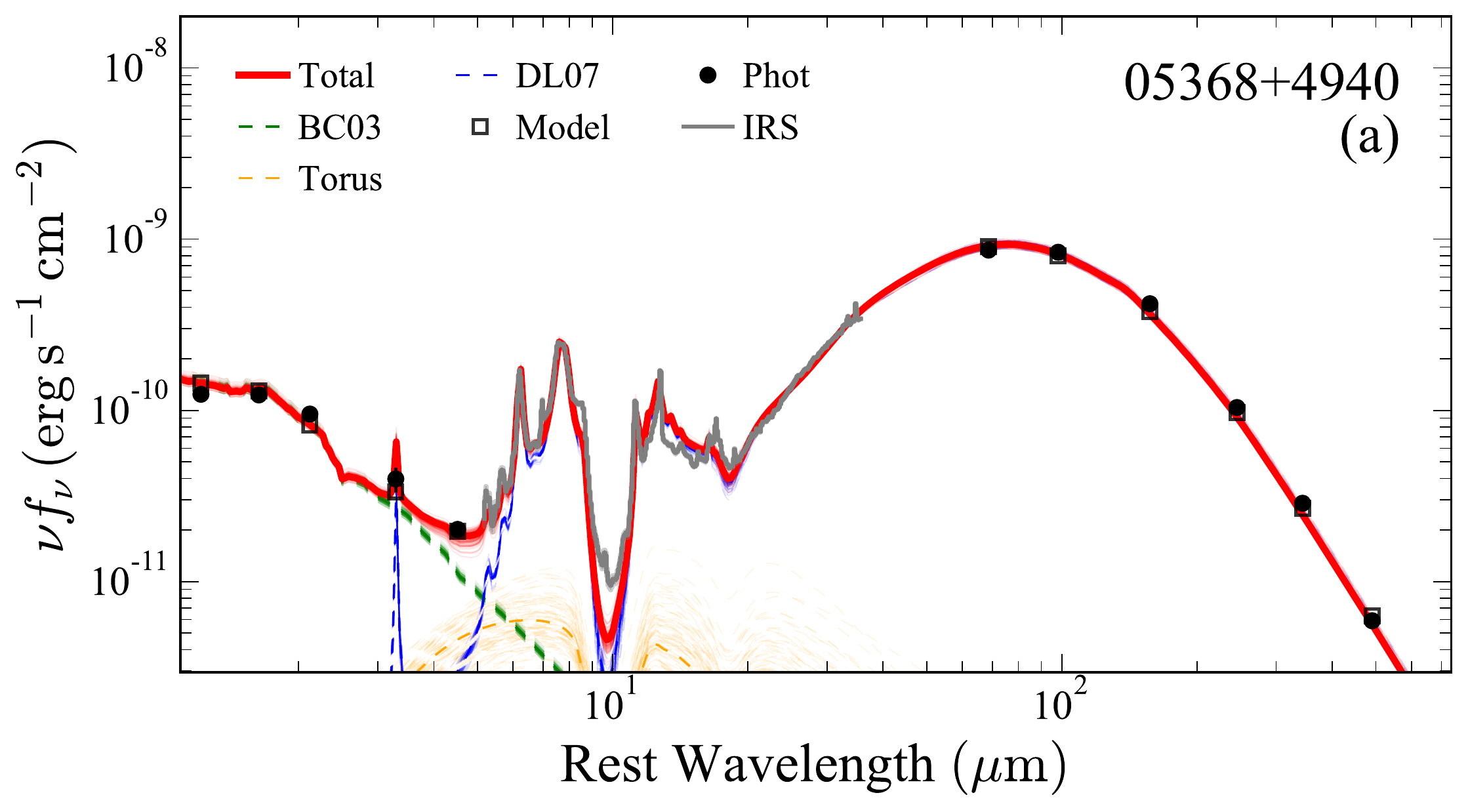} &
\includegraphics[height=0.2\textheight]{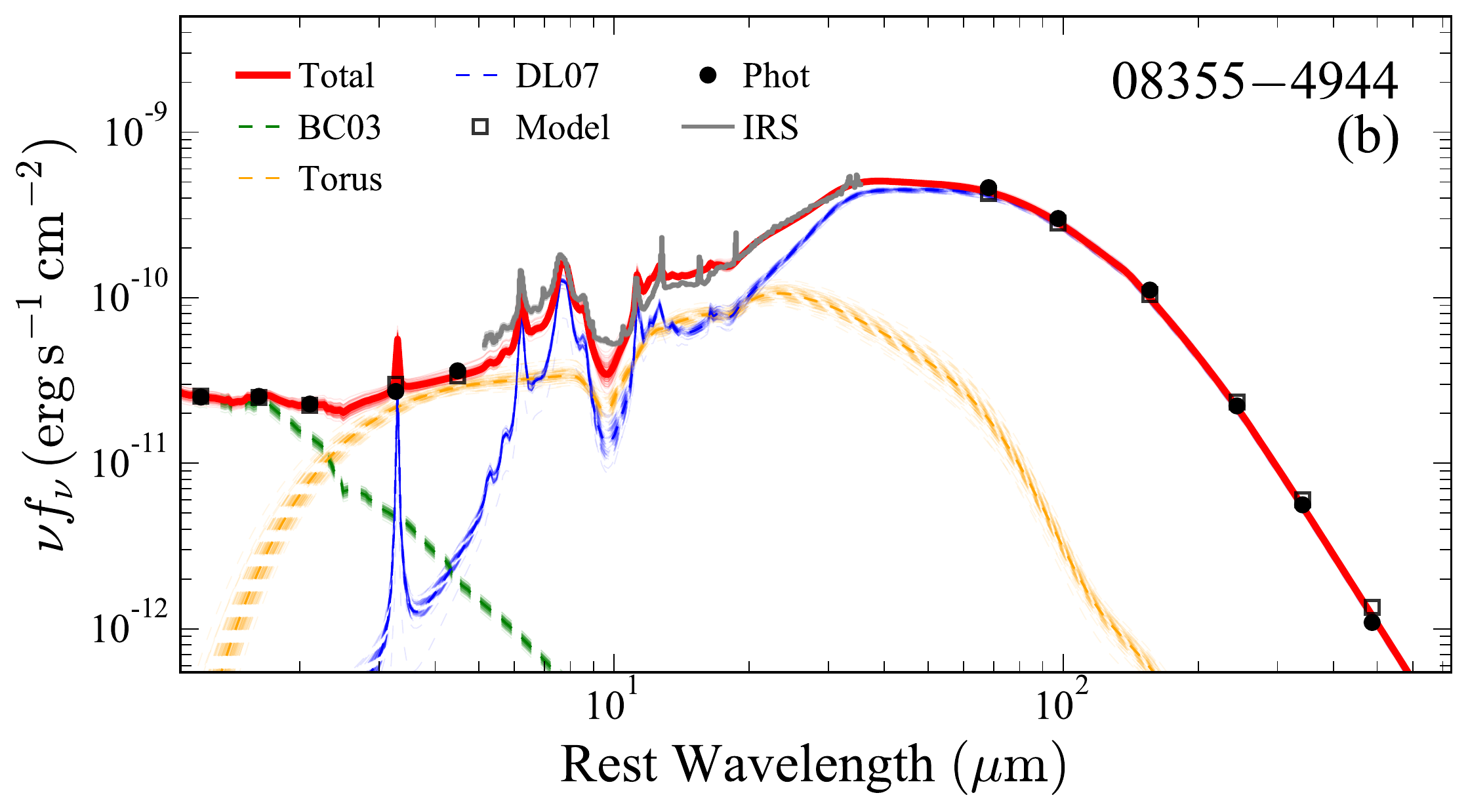} \\
\includegraphics[height=0.2\textheight]{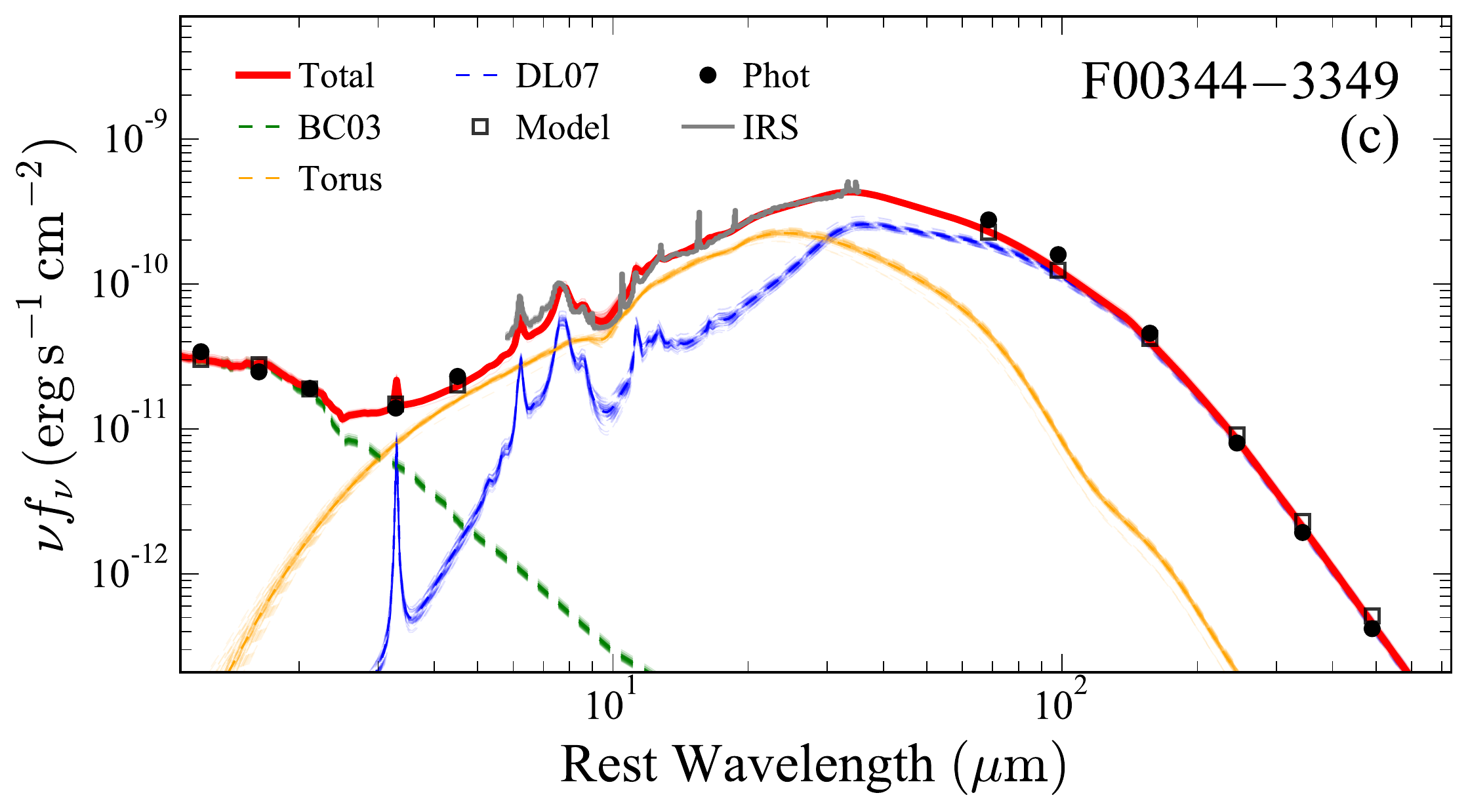} &
\includegraphics[height=0.2\textheight]{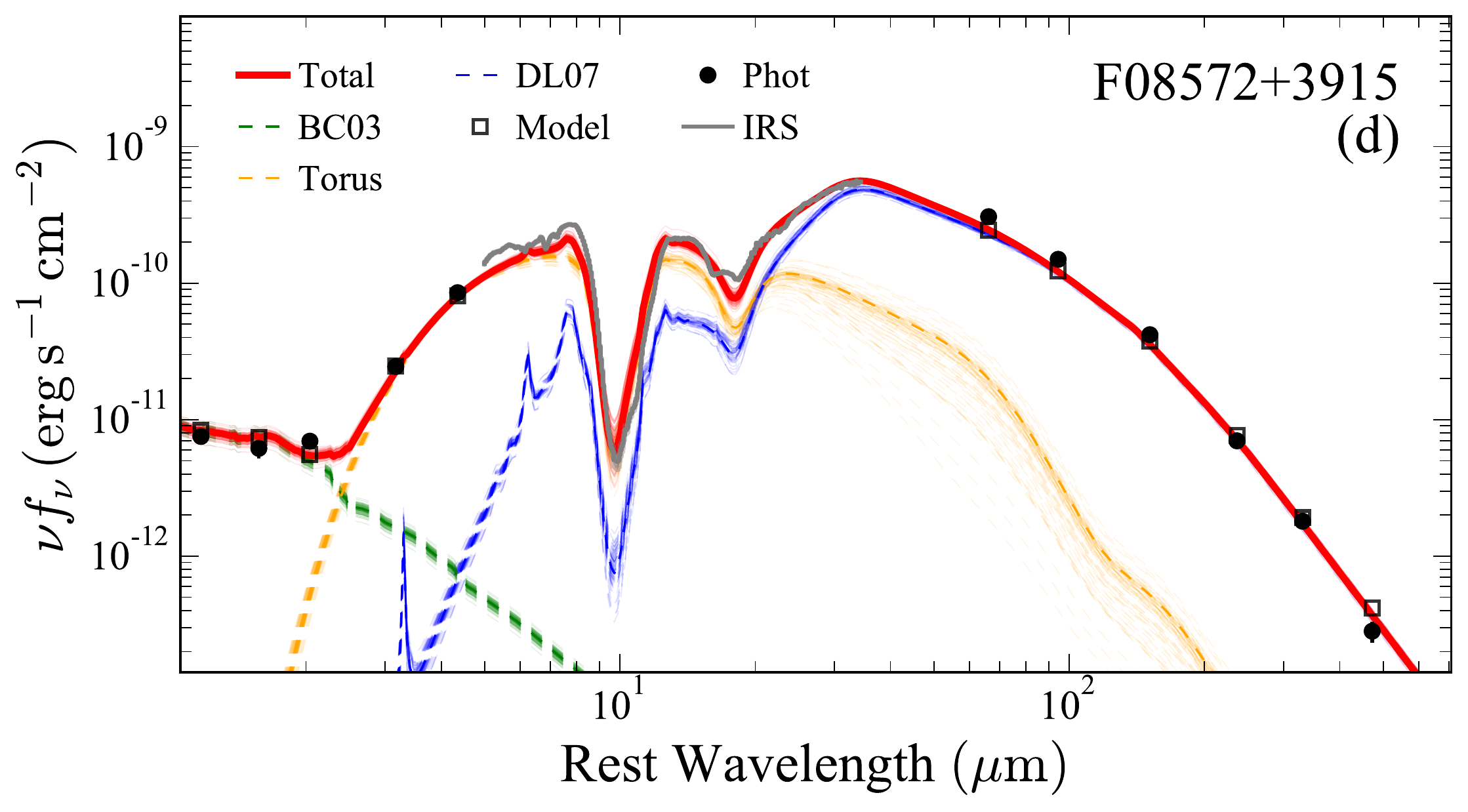} 
\end{tabular}
\caption{Examples of SED fits using IRS spectra.  The symbols are the same as 
in Figure \ref{fig:phot}.  The gray histogram plots the IRS spectrum.}
\label{fig:spec}
\end{center}
\end{figure*}

In order to test the robustness of the physical quantities derived from fitting
the photometric SED alone, we select a subsample of 61 objects whose IRS 
spectra are least affected by aperture mismatch (Appendix \ref{apd:irs}).  The 
IRS spectra provide abundant mid-IR features that allow SED fits with more 
detailed models. Unlike the fits with photometric data only, we adopt CAT3D 
torus models with a wind component (H{\"o}nig \& Kishimoto 2017).  The same 
extinction is applied to the torus and DL07 components.  In principle, the 
torus and DL07 components may suffer different levels of extinction because 
the torus resides in the nucleus while the galactic dust is distributed more 
extensively. However, allowing for separate extinctions for the torus and 
DL07 components do not significantly improve the fits.  All 9 parameters of 
the torus model, as well as \qpah\ and $\gamma$ for the DL07 component, 
are set free, in addition to the other free parameters used for the photometric 
SED fitting (Table \ref{tab:pars}).  The results are shown in Figure 
\ref{fig:spec}.  In contrast to Figure \ref{fig:phot}, data with IRS spectra 
better constrain the SED models, especially the torus and the silicate 
absorptions governed by the mid-IR extinction (see panel (b)--(d) of Figure 
\ref{fig:phot} and Figure \ref{fig:spec}).  F08572$+$3915 was not well
fit with the photometric data alone; from its IRS spectrum, we know that 
this object is highly absorbed in the mid-IR, a characteristic not revealed by 
the photometry alone.  Nevertheless, even with the addition of the IRS 
spectrum the best-fit model still lies systematically below the data at 70--160 
\micron, although \umin\ has reached the maximum.  The poor fit with the 
photometric SED likely is also due, at least in part, to this problem.

\begin{figure*}
\begin{center}
\begin{tabular}{c c}
\includegraphics[height=0.35\textheight]{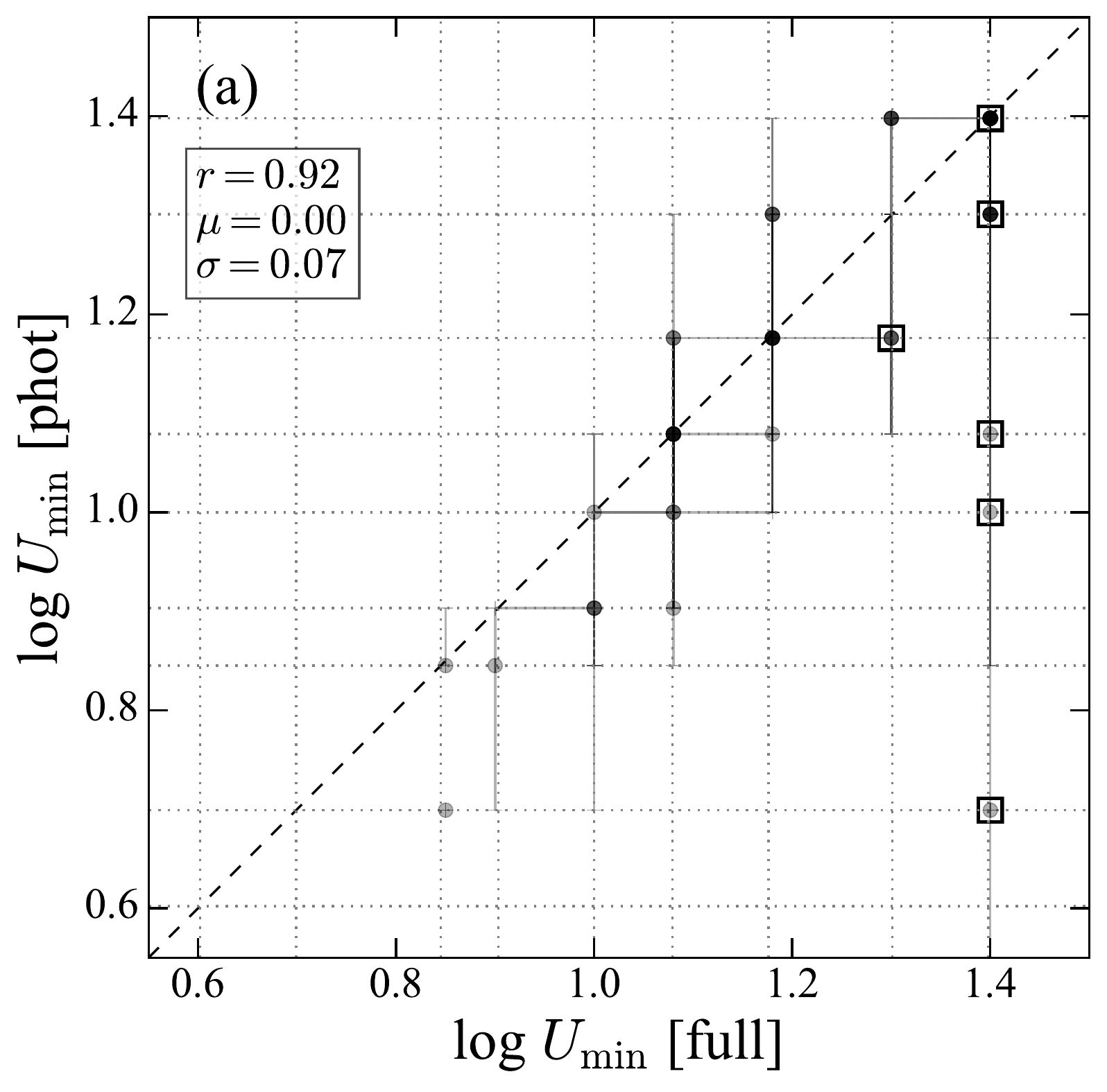} &
\includegraphics[height=0.35\textheight]{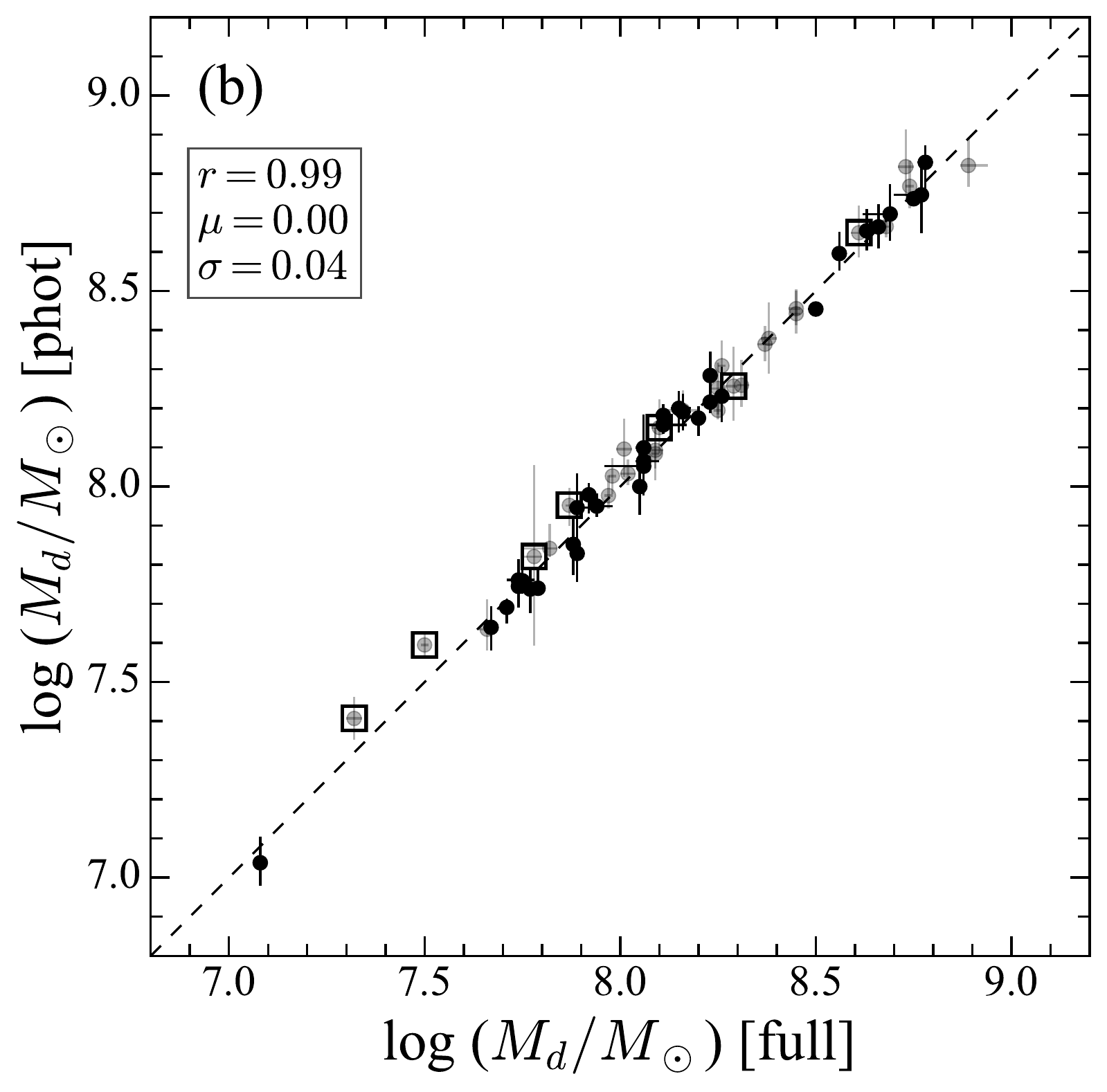}  
\end{tabular}
\caption{Comparison of the best-fit (a) \umin\ and (b) \md\ of the cold dust 
emission derived from SED fits using only photometric data versus those that 
incorporate full spectroscopic data from {\it Spitzer}/IRS.  The consistency 
is good.  Since \umin\ is a discrete parameter, the results are located 
on the dashed grids and sometimes overlap with each other; the errors are not 
resolvable if they are smaller than the grid size.  In panel (a), points with
darker shade correspond to more objects.  In panel (b), the black circles are 
objects with $\mathrm{log}\,\tau_{9.7}<0.3$, while the gray symbols are the 
rest of the objects with high mid-IR extinction.  Objects highlighted with 
boxes are those with photometric SEDs that are visually identified as having 
less robust fits (e.g., F08572$+$3915 in Figure \ref{fig:phot}d).  The 
legend in the upper-left corner shows the Pearson's correlation coefficient 
($r$) of the two quantities, and the median ($\mu$) standard deviation 
($\sigma$) of their differences ($y-x$).  The uncertain objects marked with 
boxes are excluded from the statistics. The dashed line shows the 1:1 relation.
}
\label{fig:cmp1}
\end{center}
\end{figure*}

\begin{figure*}
\begin{center}
\begin{tabular}{c c}
\includegraphics[height=0.35\textheight]{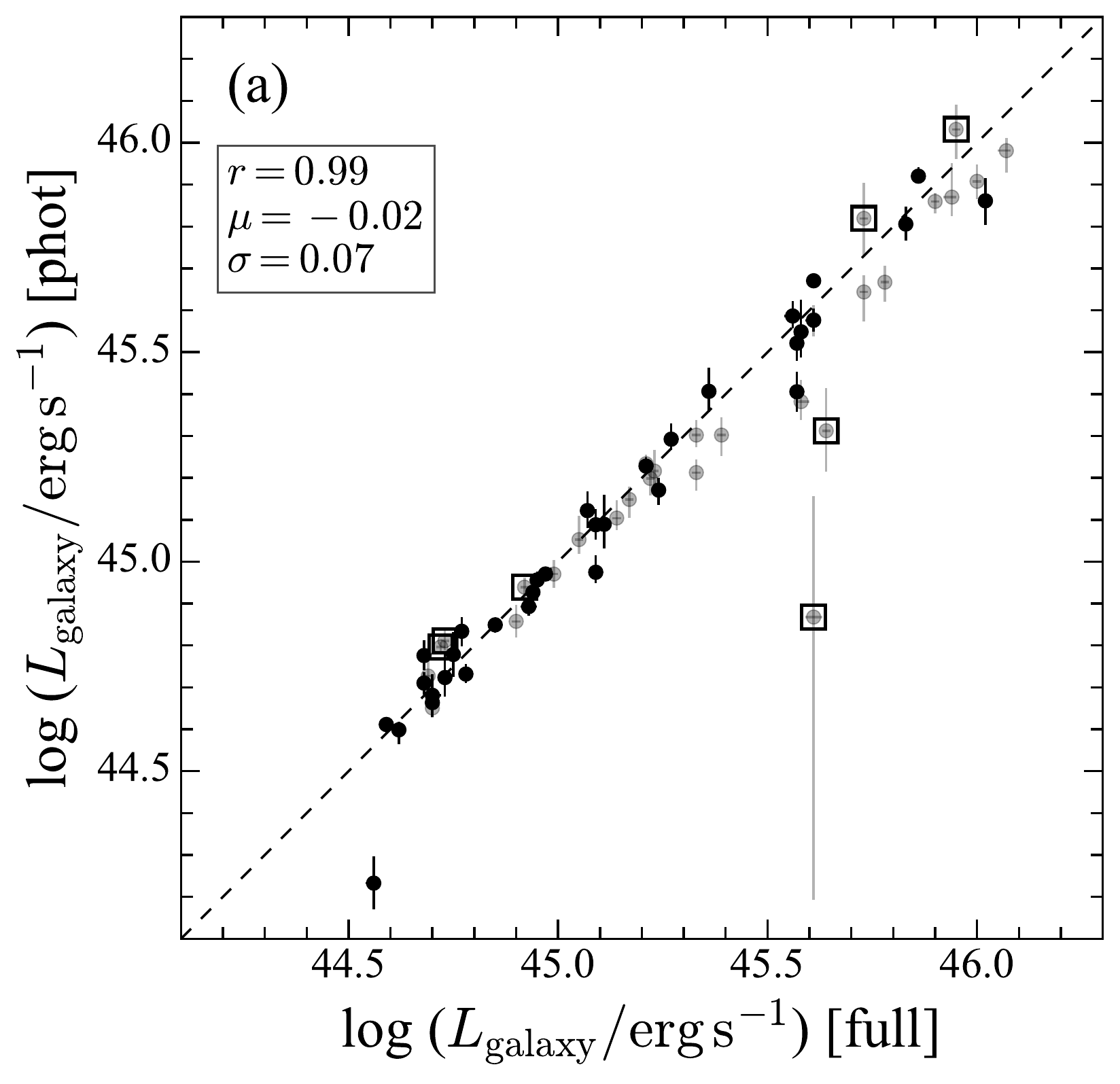}  &
\includegraphics[height=0.35\textheight]{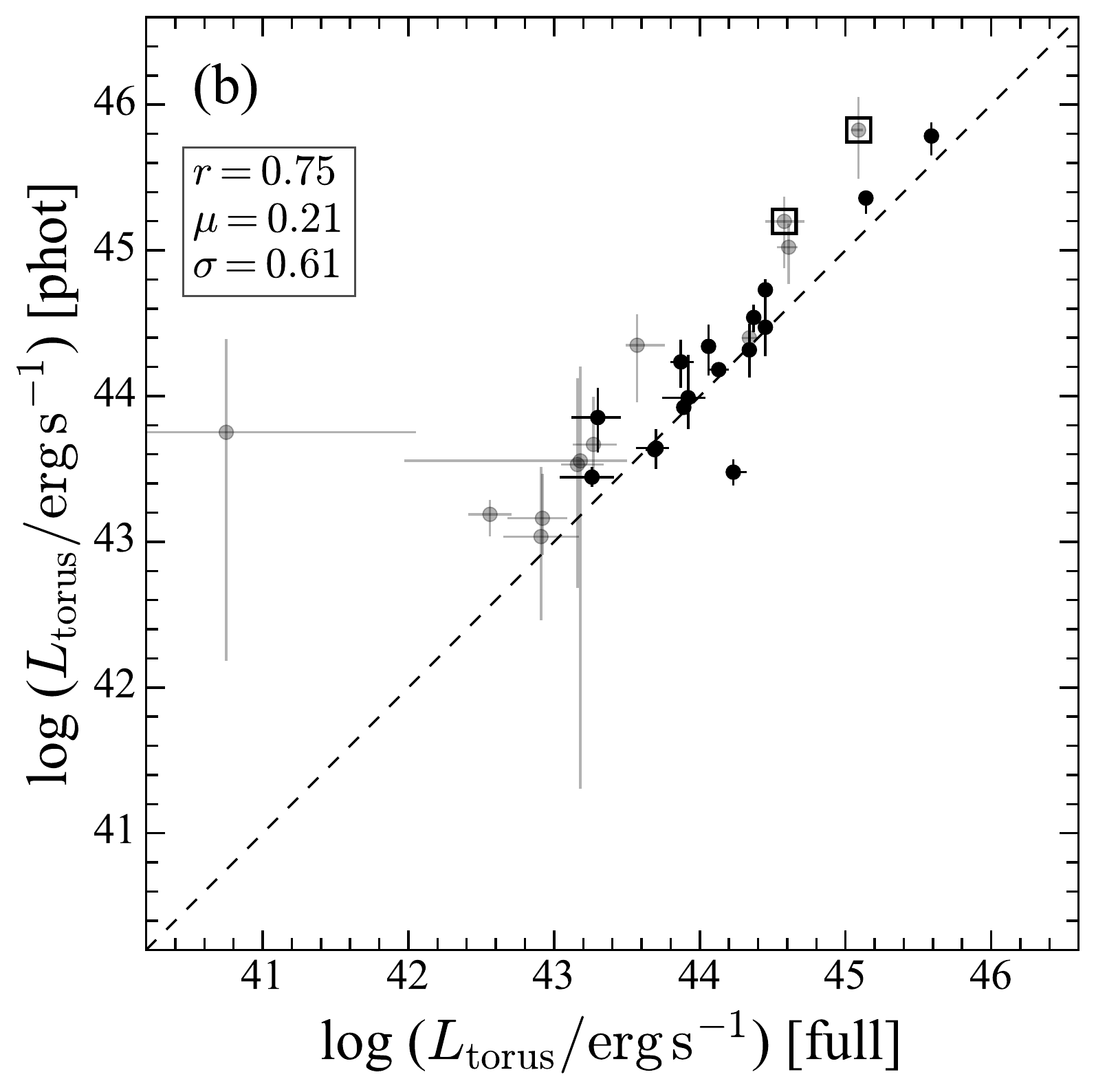}
\end{tabular}
\caption{The best-fit IR ($8-1000$ \micron) luminosity of (a) the galaxy 
(\lhost) and (b) dust torus (\ltorus) using photometric and full SEDs
are reasonably consistent with each other for the robustly fitted objects.  
The symbols are the same as in Figure \ref{fig:cmp1}.
}
\label{fig:cmp2}
\end{center}
\end{figure*}

We further confirm that, apart from the seven objects\footnote{The seven objects are 07251$-$0248,
F08572+3915, F12224$-$0624, F12243$-$0036, F13126+2453, F15250+3608, 
and F22491$-$1808.} with distinctly bad fits to the photometric SED, the best-fit 
results with and without the IRS spectra are consistent.  The two sets of fits give 
very well-matched best-fit parameters for the DL07 component (Figure \ref{fig:cmp1}).  
Objects that show less robust photometric SED fits tend to have lower \umin\ 
and \lhost\ (Figure \ref{fig:cmp2}a), likely because of overestimation of the torus 
component.  Fortunately, the dust mass is always very well matched.  For the 
case of F08572$+$3915 (Figure \ref{fig:phot}d and Figure \ref{fig:spec}), this 
is likely because \umin\ and \lhost\ decrease together.  When the torus occupies 
more of the emission from the cold dust component, it pushes the DL07 
component to peak at even longer wavelengths (lower \umin), such that given 
the same amount of emission more dust mass is required.  This effect balances
the dust mass.  As shown in Figure \ref{fig:cmp2}b, the torus luminosity and 
the fractional contribution of the torus emission to the total far-IR (8--1000 
\micron) are also reasonably constrained from the photometric SED fitting, 
albeit with $\sim 50\%$ systematic overestimation.  It is partly because 
\taumir\ is usually underestimated from the photometric SED fit unless \taumir\
is significant (e.g., $\gtrsim 2$).  Nevertheless, all of the objects showing 
significant inconsistency between the photometric and full SED fits turns out 
to be successfully identified by visual inspection of the photometric fits.

\section{Comparison between the stellar mass and star formation rate}
\label{apd:mstar}

\begin{figure*}
\begin{center}
\includegraphics[height=0.3\textheight]{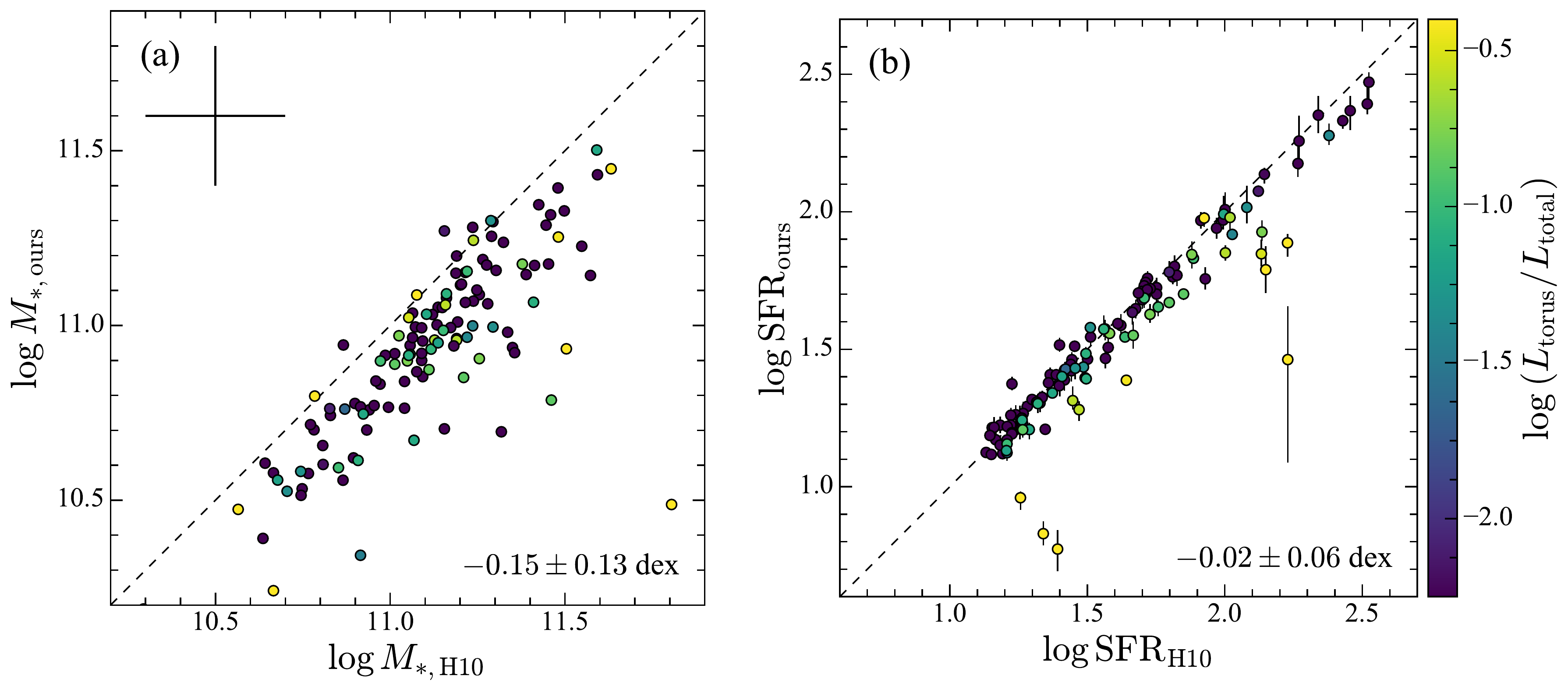} 
\caption{Comparison of our newly measured (a) stellar masses and (b) SFRs
with those provided by Howell et al. (2010; H10).  The fractional contribution 
of the torus (\ltorus) to the total IR luminosity (\ltorus+\lhost) is color 
coded with the right color bar.  The torus fraction is zero (dark purple) for 
objects fitted without a torus component.  Some objects with a high torus 
fraction tend to have \ms\ and SFR overestimated by H10, confirming that 
it is necessary to decompose the torus emission in order to robustly measure
the two quantities.
The median and standard deviation of the difference ($y-x$) for objects with 
torus fraction $\lesssim 25\%$ are shown in the lower-right corner.  The 
median uncertainty of our $M_\mathrm{*}$ measurements and a typical error bar 
of 0.2 dex from H10 are shown at the upper-left corner of (a).  The systematic 
deviation of $M_\mathrm{*}$ is not significant given the typical uncertainty;
it is likely due to the choice of mass-to-light ratio.  The two sets of SFRs 
are in good agreement.
}
\label{fig:h10}
\end{center}
\end{figure*}

The stellar masses and SFRs for a large fraction of the GOALS galaxies were 
also derived by Howell et al. (2010; their Table 3).  They estimated the 
integrated stellar mass using 2MASS $K_s$-band luminosity, supplemented with 
{\it Spitzer}/IRAC 3.6 \micron\ photometry.  They assumed a $M/L$ according to 
Lacey et al. (2008), who assumed an IMF close to that of Kroupa (2001) for
quiescent galaxies and a top-heavy IMF for starburst galaxies.  Since it is 
not straightforward to convert their IMFs to our choice of Chabrier (2003)
IMF, we directly compare our newly derived \ms\ with Howell et al.'s results 
(Figure \ref{fig:h10}a).  Howell et al. did not consider AGN torus emission, 
which may significantly contaminate the $K_s$ and IRAC 3.6 \micron\ bands.
We use the fraction of torus emission derived from our analysis to evaluate 
the possible bias of \ms.  Objects with high torus fractions tend to exhibit 
the largest systematic deviations between the two sets of measurements.
Excluding the objects with the most significant torus emission 
($\log\,L_\mathrm{torus}/L_\mathrm{total}>-0.6$), our stellar masses are still 
systematically lower than those of Howell et al. by $0.15 \pm 0.13$ dex.  This 
is likely due to the different choices of $M/L$, including the effect of 
different IMFs.  Nevertheless, considering the typical 0.2 dex uncertainty of 
stellar masses (Conroy 2013), this level of discrepancy is not serious.  In 
fact, our stellar masses are on average $0.15 \pm 0.31$ dex 
higher than those of U et al. (2012), in spite of the same (Chabrier) IMF used 
in both.  Further detailed study of the full UV-to-IR SED is necessary to derive 
more robust stellar masses, but this is outside of the scope of the current work.

Howell et al. (2010) calculated SFRs using the formalism of Kennicutt (1998b) 
that combines UV and IR emission.  We divide their SFRs by a factor of 1.5 
to convert their scale based on the Salpeter IMF to that of the Chabrier IMF.
As Howell et al. caution, some of their SFRs are severely affected by AGN 
contamination (Figure \ref{fig:h10}b).  Excluding the objects with torus 
fractions $\gtrsim$ 25\%, we find that Howell et al.'s SFRs are consistent 
with ours.

\section{Dust masses from the modified blackbody model}
\label{apd:mbb}

The templates provided by DL07 are limited to $U_\mathrm{min} \leq 25$.  Some 
of the LIRGs in our sample saturate at this limit, and their fits can still be 
improved.  The dust masses may be overestimated in these objects.  This is a 
non-trivial point, in light of our conclusion that late-stage mergers have 
higher gas mass fractions than the earlier stages (Figure \ref{fig:fgas}).  As 
\umin\ tends to increase toward later stage mergers (Figure \ref{fig:umin}), 
their apparently higher gas mass fractions may be an artifact.  In order to 
quantify this possible bias on the dust mass and test the robustness of the 
gas fraction distribution, we obtain an alternative estimate of the dust mass 
by fitting the far-IR (100 to 500 \micron) SED with the MBB model: 

\begin{equation}
f_\mathrm{\nu, MBB} = \frac{(1+z)^2\,M_{d}\,\kappa_\mathrm{abs}(\lambda_0)\,
(\lambda_0/\lambda)^\beta\,B_\nu(T_d)}{D_{L}^2},
\end{equation}

\noindent
where $f_\mathrm{\nu, MBB}$ is the rest-frame flux density, $D_L$ is the
luminosity distance, $z$ is the redshift, and $B_\nu(T_d)$ is the Planck 
function with dust temperature $T_d$.  Following Bianchi (2013), we adopt the 
grain absorption cross-section per unit mass at 250 \micron,
$\kappa_\mathrm{abs}(250\,\micron) = 4.0\,\mathrm{cm^2 g^{-1}}$, and fix 
$\beta = 2.08$.  Our method is adapted from Shangguan et al. (2018), who 
provide a detailed comparison between the MBB and DL07 models.  

\begin{figure*}
\begin{center}
\includegraphics[height=0.3\textheight]{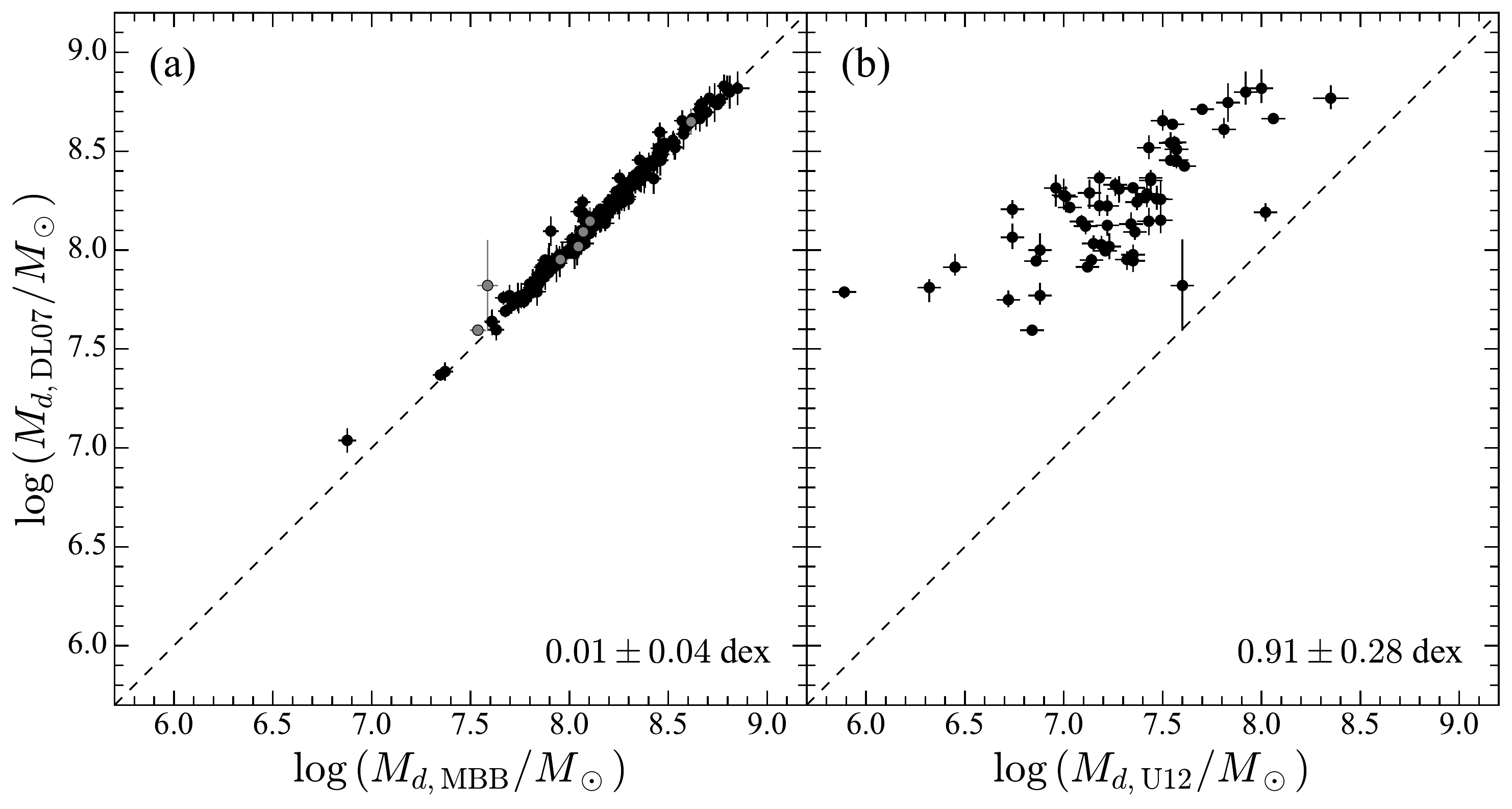} 
\caption{Comparison of our dust masses derived from DL07 models fit to the 
100--500 \micron\ SED with (a) dust masses obtained from MBB models fit to the 
same data and (b) dust masses published by U et al. (2012; U12).  The consistency 
with the MBB-derived dust masses is very good, with median deviation 0.01 dex 
and standard deviation 0.04 dex.  The gray circles are objects less robustly 
fitted with the DL07 model.  By comparison, the dust masses of U12 are significantly 
lower ($0.91\pm0.28$ dex) than ours.
}
\label{fig:mbb1}
\end{center}
\end{figure*}

\begin{figure*}
\begin{center}
\includegraphics[height=0.3\textheight]{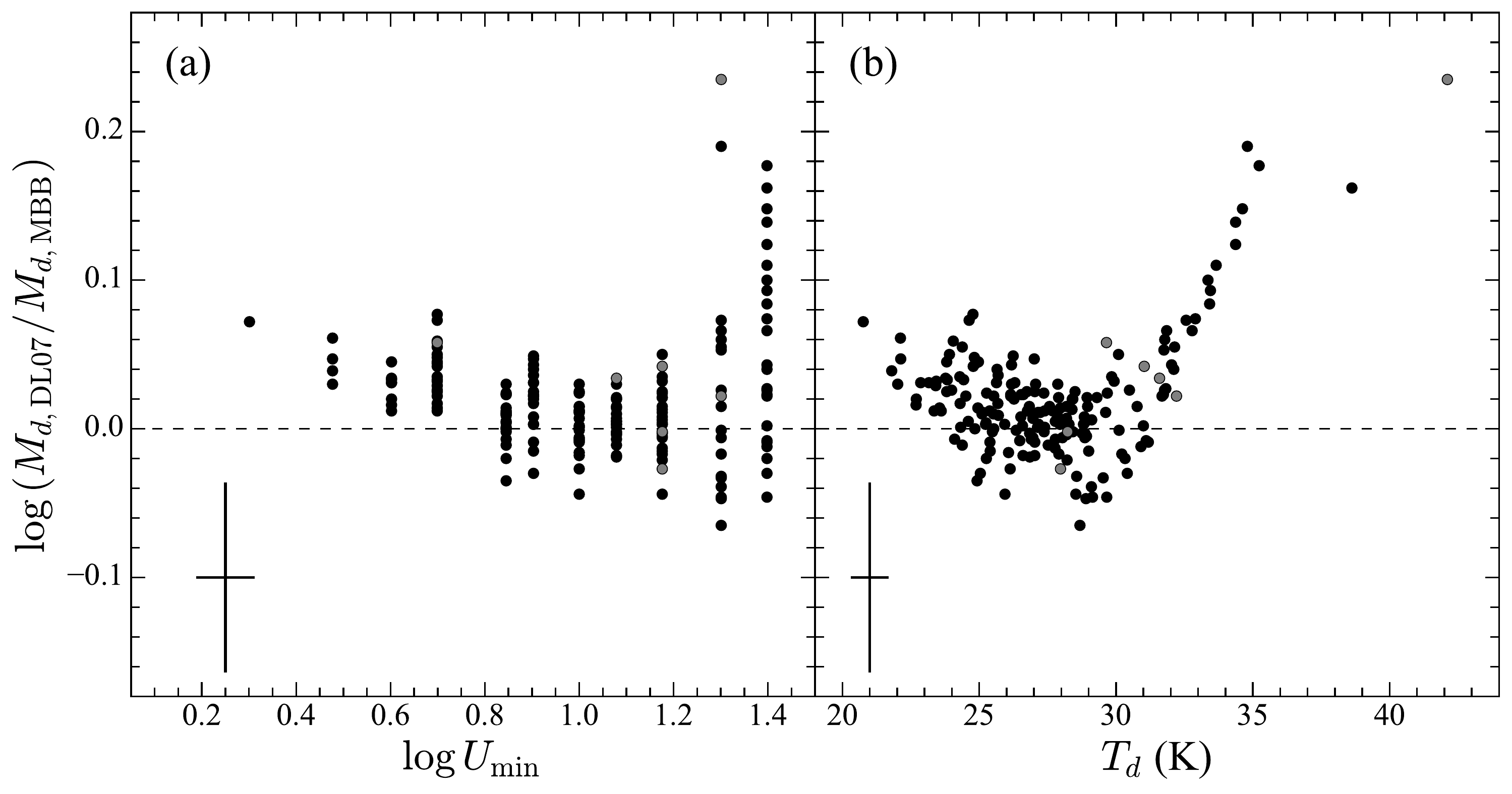} 
\caption{The deviation of the dust mass (between DL07 and MBB) is plotted 
against (a) \umin\ and (b) $T_d$.  For objects with $U_\mathrm{min} < 25$ or 
$T_d \lesssim 33$ K, the deviation is within $\sim$0.1 dex.  However, at 
$U_\mathrm{min} = 25$ or $T_d > 33$ K, the DL07 dust mass is systematically 
higher than the MBB dust mass, which is exactly as expected because the DL07
templates are limited by $U_\mathrm{min} \leq 25$.  However, even the 
highest deviation ($\sim 0.2$ dex) is still moderate.
}
\label{fig:mbb2}
\end{center}
\end{figure*}

As Figure \ref{fig:mbb1}a shows, the dust masses derived from the two methods 
agree with each other extremely well for the entire sample.  The median 
deviation is 0.01 dex, and the standard deviation is 0.04 dex.  The deviation 
indeed increases as a function of \umin\ or $T_d$ (Figure \ref{fig:mbb2}), but 
hardly beyond 0.2 dex.  F08572+3915, whose torus emission and mid-IR extinction 
are both very strong, is the only object whose dust mass deviates by more than 
0.2 dex between the two methods.  The poor quality of its SED fit can be 
readily identified.  In sharp contrast, the dust masses of U et al. (2012; U12) are 
systematically and severely understimated, by $\sim 0.9$ dex (Figure 
\ref{fig:mbb1}b).  U12 derived dust masses by fitting the mid-IR to far-IR SED with 
a truncated power-law plus an MBB models (Casey 2012).  The main reason that 
U12 systematically underestimated their dust masses is probably because they 
adopted a too large value of the grain absorption cross-section per unit mass.  
U12 quote $\kappa_\mathrm{abs}(850\,\micron) = 0.15\,\mathrm{m^{2}\,kg^{-1}}$ 
referring to Weingartner \& Draine (2001) and Dunne et al. (2003).  Weingartner \& 
Draine (2001) provide a much lower value of 
$\kappa_\mathrm{abs}(850\,\micron)=0.0383\,\mathrm{m^{2}\,kg^{-1}}$ (Draine 2003), 
while $0.15\,\mathrm{m^{2}\,kg^{-1}}$ is the upper limit for the diffuse ISM of 
extragalactic systems quoted by Dunne et al. (2003).  This could account for $\sim 0.6$ 
dex of the deviation.  Moreover, the dust temperature derived by U12 is also systematically 
higher than ours (median difference $3.6 \pm 4.3$ K), likely due to their simplified 
modeling.  This may also contribute to the systematic deviation in dust mass.

\begin{figure*}
\begin{center}
\includegraphics[height=0.3\textheight]{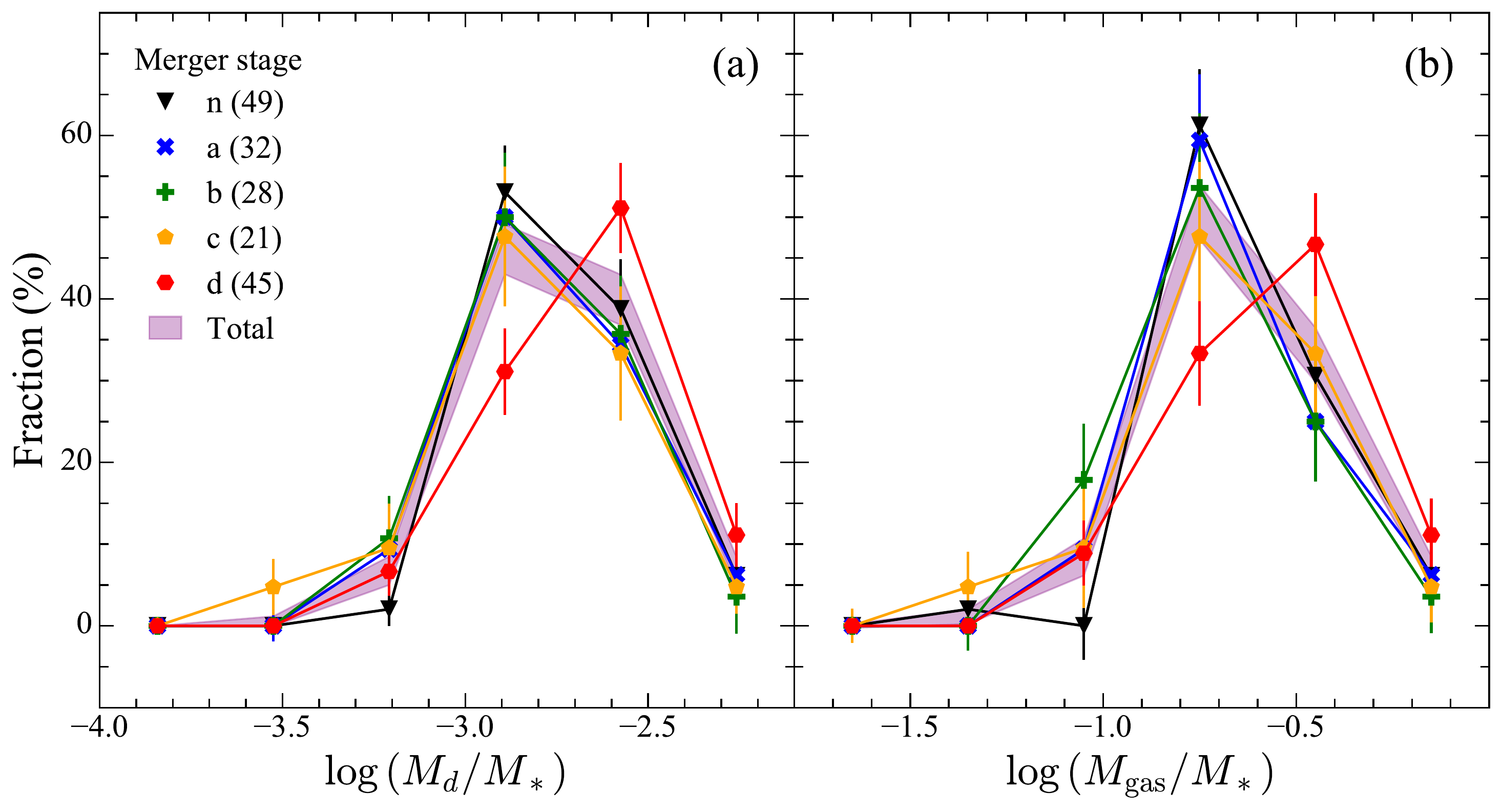} 
\caption{Same as Figure \ref{fig:fgas}, except that the dust (and gas) 
masses are based on the MBB model.  There is not much difference between the 
distributions based on the two sets of models.
}
\label{fig:fgas2}
\end{center}
\end{figure*}

Figure \ref{fig:fgas2} reexamines the results presented in Figure \ref{fig:fgas}
concerning the distributions of dust and gas mass fractions as a function of 
galaxy merger stage, using dust masses derived from MBB fits instead of DL07 
models.  The two sets of results are indistinguishable.

\begin{longrotatetable}
\begin{deluxetable*}{l l r c c r r r c c c c c c c}
\tabletypesize{\scriptsize}
\tablecaption{Physical Properties of the Sample \label{tab:sample}}
\tablehead{
\colhead{Name} &
\colhead{$z$} &
\colhead{$D_L$} &
\colhead{Merger} &
\colhead{AGN} &
\colhead{References} &
\colhead{$\mathrm{log}\,M_*$} &
\colhead{$\mathrm{log}\,\tau_{9.7}$} &
\colhead{$\mathrm{log}\,L_\mathrm{torus}$} &
\colhead{$\mathrm{log}\,L_\mathrm{galaxy}$} &
\colhead{$\mathrm{log}\,U_\mathrm{min}$} &
\colhead{$\mathrm{log}\,M_d$} &
\colhead{$\mathrm{log}\,\delta_\mathrm{GDR}$} &
\colhead{$\mathrm{log}\,M_\mathrm{gas}$} &
\colhead{$\mathrm{log}\,\mathrm{SFR}$} \\
 &
 &
\colhead{(Mpc)} &
\colhead{Stage} &
 &
 &
\colhead{($M_\odot$)} &
 &
\colhead{($\mathrm{erg\,s^{-1}}$)} &
\colhead{($\mathrm{erg\,s^{-1}}$)} &
 &
\colhead{($M_\odot$)} &
 &
\colhead{($M_\odot$)} &
\colhead{($M_\odot\,\mathrm{yr^{-1}}$)} \\
\colhead{(1)} &
\colhead{(2)} &
\colhead{(3)} &
\colhead{(4)} &
\colhead{(5)} &
\colhead{(6)} &
\colhead{(7)} &
\colhead{(8)} &
\colhead{(9)} &
\colhead{(10)} &
\colhead{(11)} &
\colhead{(12)} &
\colhead{(13)} &
\colhead{(14)} &
\colhead{(15)}
}
\startdata
 F00073$+$2538        & 0.0152 &  67.2 &        b &       N &                  23 & $11.12\pm0.20$ &  $-0.16^{+0.32}_{-0.41}$ &                 \nodata & $44.64^{+0.01}_{-0.01}$ & $0.85^{+0.06}_{-0.00}$ & $7.96^{+0.02}_{-0.02}$ & 2.08 & $10.04\pm0.20$ & $1.12^{+0.02}_{-0.01}$ \\
 F00085$-$1223        & 0.0196 &  86.7 &        d &       Y &              20, 23 & $10.76\pm0.20$ &  $-1.10^{+1.39}_{-1.93}$ &                 \nodata & $45.05^{+0.04}_{-0.04}$ & $1.30^{+0.10}_{-0.12}$ & $7.77^{+0.07}_{-0.05}$ & 2.10 & $ 9.86\pm0.20$ & $1.47^{+0.04}_{-0.04}$ \\
 F00163$-$1039        & 0.0273 & 121.1 &        b &       N &                  23 & $10.91\pm0.20$ &  $-2.37^{+1.32}_{-1.17}$ & $44.28^{+0.03}_{-0.04}$ & $44.97^{+0.03}_{-0.02}$ & $1.00^{+0.08}_{-0.00}$ & $8.12^{+0.03}_{-0.04}$ & 2.09 & $10.25\pm0.21$ & $1.56^{+0.02}_{-0.02}$ \\
 F00344$-$3349        & 0.0206 &  91.8 &        c &       N &                  23 & $10.24\pm0.20$ &  $ 0.47^{+0.15}_{-0.34}$ & $44.73^{+0.07}_{-0.11}$ & $44.23^{+0.06}_{-0.06}$ & $1.40^{+0.00}_{-0.10}$ & $7.04^{+0.07}_{-0.06}$ & 2.17 & $ 9.26\pm0.21$ & $0.77^{+0.07}_{-0.08}$ \\
 F00402$-$2349        & 0.0226 & 100.3 &        b &       Y &              18, 23 & $11.30\pm0.20$ &  $-2.96^{+1.22}_{-0.72}$ & $43.81^{+0.06}_{-0.04}$ & $45.10^{+0.01}_{-0.02}$ & $1.00^{+0.00}_{-0.00}$ & $8.27^{+0.02}_{-0.01}$ & 2.09 & $10.35\pm0.20$ & $1.58^{+0.02}_{-0.01}$ \\
 F00506$+$7248        & 0.0157 &  72.0 &        c &       N &                   2 & $10.61\pm0.20$ &  $-2.24^{+1.30}_{-1.26}$ & $44.05^{+0.13}_{-0.27}$ & $45.06^{+0.06}_{-0.07}$ & $1.18^{+0.22}_{-0.10}$ & $7.98^{+0.08}_{-0.08}$ & 2.11 & $10.16\pm0.21$ & $1.57^{+0.05}_{-0.03}$ \\
 F00548$+$4331        & 0.0181 &  80.2 &        a & \nodata &             \nodata & $11.04\pm0.20$ &  $-1.51^{+1.42}_{-1.77}$ &                 \nodata & $44.77^{+0.03}_{-0.02}$ & $1.00^{+0.08}_{-0.00}$ & $7.89^{+0.04}_{-0.04}$ & 2.08 & $ 9.95\pm0.24$ & $1.22^{+0.07}_{-0.05}$ \\
 F01053$-$1746        & 0.0201 &  88.2 &        c &       N &                  23 & $10.85\pm0.20$ &  $-0.57^{+0.28}_{-0.51}$ & $44.39^{+0.03}_{-0.03}$ & $45.19^{+0.02}_{-0.02}$ & $1.08^{+0.00}_{-0.00}$ & $8.27^{+0.02}_{-0.02}$ & 2.09 & $10.37\pm0.20$ & $1.67^{+0.02}_{-0.02}$ \\
 F01076$-$1707        & 0.0351 & 156.5 &        n &       N &                  23 & $11.15\pm0.20$ &  $-1.35^{+1.37}_{-1.76}$ &                 \nodata & $45.29^{+0.04}_{-0.03}$ & $1.18^{+0.12}_{-0.10}$ & $8.28^{+0.06}_{-0.07}$ & 2.08 & $10.37\pm0.21$ & $1.76^{+0.03}_{-0.02}$ \\
 F01159$-$4443        & 0.0229 & 102.7 &        b &       N &                  23 & $11.05\pm0.20$ &  $-0.94^{+0.98}_{-1.89}$ &                 \nodata & $45.03^{+0.02}_{-0.02}$ & $1.08^{+0.10}_{-0.08}$ & $7.93^{+0.05}_{-0.05}$ & 2.08 & $10.01\pm0.21$ & $1.51^{+0.02}_{-0.02}$ \\
 F01173$+$1405        & 0.0312 & 137.9 &        b &       N &                  23 & $10.59\pm0.20$ &  $-0.37^{+0.54}_{-1.07}$ & $44.23^{+0.15}_{-0.18}$ & $45.17^{+0.03}_{-0.04}$ & $1.40^{+0.00}_{-0.00}$ & $7.95^{+0.03}_{-0.03}$ & 2.12 & $10.07\pm0.20$ & $1.65^{+0.02}_{-0.03}$ \\
 F01325$-$3623        & 0.0159 &  70.6 &        n & \nodata &             \nodata & $10.75\pm0.20$ &  $-1.48^{+1.47}_{-1.72}$ &                 \nodata & $44.66^{+0.04}_{-0.03}$ & $0.85^{+0.06}_{-0.15}$ & $8.10^{+0.09}_{-0.06}$ & 2.10 & $10.19\pm0.21$ & $1.14^{+0.03}_{-0.04}$ \\
 F01341$-$3735        & 0.0173 &  76.9 &        a &       N &                  23 & $10.87\pm0.20$ &  $-0.93^{+1.04}_{-1.95}$ &                 \nodata & $44.74^{+0.02}_{-0.02}$ & $0.90^{+0.10}_{-0.06}$ & $7.94^{+0.04}_{-0.05}$ & 2.09 & $10.04\pm0.20$ & $1.23^{+0.02}_{-0.02}$ \\
 F01364$-$1042\tnm{a} & 0.0483 & 216.9 &        d &       N &                  23 & $10.39\pm0.22$ &  $ 1.26^{+0.04}_{-0.05}$ &                 \nodata & $45.51^{+0.03}_{-0.04}$ & $1.30^{+0.00}_{-0.22}$ & $8.09^{+0.05}_{-0.04}$ & 2.15 & $10.23\pm0.20$ & $1.97^{+0.03}_{-0.03}$ \\
 F01417$+$1651        & 0.0274 & 120.3 &        a &       Y &                  23 & $10.42\pm0.20$ &  $ 0.35^{+0.60}_{-2.29}$ &                 \nodata & $45.12^{+0.07}_{-0.05}$ & $1.30^{+0.10}_{-0.12}$ & $7.95^{+0.07}_{-0.06}$ & 2.14 & $10.09\pm0.21$ & $1.59^{+0.06}_{-0.05}$ \\
\enddata
\tablecomments{
(1) Source name.
(2) Redshift from NASA/IPAC Extragalactic Database (NED).
(3)  The luminosity distance in Mpc derived by correcting the heliocentric
velocity for the 3-attractor flow model of Mould et al. (2000) using
$\Omega_m=0.308$, $\Omega_\Lambda=0.692$, and
$H_0=67.8\,\mathrm{km\,s^{-1}\,Mpc^{-1}}$ (Planck Collaboration et al. 2016).
(4) The merger stage adopted from Stierwalt et al (2013): n = nonmerger, a =
pre-merger, b = early-stage merger, c = mid-stage merger, and d = late-stage
merger.  Ten objects marked as ``?'' are not included in Stierwalt et al. (2013).
(5) Type of nuclear activity.
(6) References for nuclear activity.
(7) Stellar mass derived from optical color and $J$-band absolute magnitude
(Bell \& de Jong 2001), converted to Chabrier (2003) IMF.
(8) The optical depth at 9.7 \micron.
(9) The IR (8--1000 \micron) luminosity of the torus derived from the SED
fitting.
(10) The IR (8--1000 \micron) luminosity of the host galaxy cold dust emission
derived from the SED fitting.  The best-fit extinction model is applied when
calculating both \ltorus\ and \lhost.
(11)  Best-fit minimum intensity of the interstellar radiation field relative to
that measured in the solar neighborhood.  The quoted uncertainties represent the
68\% confidence interval determined from the 16th and 84th percentile of the
marginalized posterior probability density function.  However, for some
objects, the  probability density function is not well resolved, and the
lower (or upper) uncertainty is reported as ``0.00''.
(12) Best-fit total dust mass.
(13) Gas-to-dust ratio estimated in Section \ref{ssec:mass}.
(14) Total gas mass including helium and heavier elements.
(15) The SFR calculated from \lhost\ (Col. 10) using the Equation (4) of
Kennicutt (1998b) converted to the Chabrier (2003) IMF by dividing by a factor of
1.5.
\\
References: (1) Albrecht et al. (2007); (2) Alonso-Herrero et al. (2009);
(3) Alonso-Herrero et al. (2012); (4) Baan et al. (1998); (5) Corbett et al.
(2003); (6) Farrah et al. (2007); (7) Gon{\c c}alves et al. (1999); (8) Ho et
al. (1997); (9) Imanishi (2006); (10) Iwasawa et al. (2011); (11) Inami et al.
(2013); (12) Kinney et al. (1993); (13) Koss et al. (2013); (14) L{\'{\i}}pari
et al. (2000); (15) Masetti et al. (2008); (16) Nardini et al. (2010); (17)
Ohyama et al. (2015); (18) Petric et al. (2011); (19) Ricci et al. (2016);
(20) Ricci et al. (2017); (21) Tueller et al. (2008); (22) Torres-Alb{\`a}
et al.(2018); (23) Yuan et al. (2010); (24) Zink et al. (2000).
\\
(This table is available in its entirety in machine-readable form.)
}
\tablenotetext{a}{The fitting is not robust from the visual inspection.}
\end{deluxetable*}
\end{longrotatetable}

\begin{table*}
  \begin{center}
  \caption{Model Parameters and Priors}
  \label{tab:pars}
  \begin{tabular}{ccccl}
    \hline
    \hline
    Models                       & Parameters       & Units                  & Discreteness & \mcl{1}{c}{Priors}                   \\
    (1)                          & (2)              & (3)                    & (4)          & \mcl{1}{c}{(5)}                      \\\hline
    \multirow{2}{*}{BC03}        & $M_*$            & $M_\odot$              & \ding{56}    & [$10^6$, $10^{14}$]                  \\
                                 & $t$              & Gyr                    & \ding{52}    & 5 (fixed)                            \\\hline
                    Extinction   & \taumir          & --                     & \ding{52}    & [$10^{-4}$, $10^{1.5}$]              \\\hline
                                 & $i$              & --                     & \ding{52}    & [0.0, 90.0]                          \\
                                 & $a$              & --                     & \ding{52}    & [-2.5, -0.25] or [-3.0, -0.5]        \\
                    CAT3D        & $N_0$            & --                     & \ding{52}    & [5.0, 10.0]                          \\
                    (Basic)      & $h$              & --                     & \ding{52}    & [0.25, 1.5] or [0.1, 0.5]            \\
                                 & $L$              & $\mathrm{erg\:s^{-1}}$ & \ding{56}    & [$10^{38}$, $10^{48}$]               \\\cline{2-5}
    \multirow{4}{*}{(Wind)}      & $f_\mathrm{wd}$  & --                     & \ding{52}    & [0.15, 1.75]                         \\
                                 & $a_w$            & --                     & \ding{52}    & [-2.5,  -0.5]                        \\
                                 & $\theta_w$       & --                     & \ding{52}    & [30, 45]                             \\
                                 & $\sigma_\theta$  & --                     & \ding{52}    & [7.5, 15]                            \\\hline
    \multirow{6}{*}{DL07}        & $U_\mathrm{min}$ & --                     & \ding{52}    & [0.10, 25.0]                         \\
                                 & $U_\mathrm{max}$ & --                     & \ding{52}    & $10^6$ (fixed)                       \\
                                 & $\alpha$         & --                     & \ding{56}    & 2 (fixed)                            \\
                                 & $q_\mathrm{PAH}$ & --                     & \ding{52}    & 0.47 (fixed) or [0.3, 5.0]           \\
                                 & $\gamma$         & --                     & \ding{56}    & 0.03 (fixed) or [0.0, 1.0]           \\
                                 & $M_d$            & $M_\odot$              & \ding{56}    & [$10^5$, $10^{11}$]                  \\\hline
  \end{tabular}
  \end{center}
\tablecomments{
(1) Model components.
(2) The parameters of each model.
(3) The units of the parameters.
(4) Whether the parameter is discrete and requires interpolation to implement
the MCMC fitting.
(5) The range of the priors.  For parameters with two priors, the first is for
the fits with photometric SEDs, while the second is for the fits with the full
SEDs.
}
\end{table*}

\begin{deluxetable*}{l l l r c l l l}
\tabletypesize{\scriptsize}
\tablecaption{Objects with \spitzer/IRS spectra \label{tab:irs}}
\tablehead{
\colhead{Name} &
\colhead{$f_\mathrm{SL}$} &
\colhead{$f_\mathrm{IRS}$} &
\colhead{$\mathrm{log}\,\tau_{9.7}$} &
\colhead{$\mathrm{log}\,U_\mathrm{min}$} &
\colhead{$\mathrm{log}\,M_d$} &
\colhead{$\mathrm{log}\,L_\mathrm{torus}$} &
\colhead{$\mathrm{log}\,L_\mathrm{galaxy}$}
\\
 &
 &
 &
 &
 &
\colhead{($M_\odot$)} &
\colhead{($\mathrm{erg\,s^{-1}}$)} &
\colhead{($\mathrm{erg\,s^{-1}}$)}
\\
\colhead{(1)} &
\colhead{(2)} &
\colhead{(3)} &
\colhead{(4)} &
\colhead{(5)} &
\colhead{(6)} &
\colhead{(7)} &
\colhead{(8)}
}
\startdata
     F00344$-$3349 & 1.15 & 1.14 & $-0.14^{+0.05}_{-0.06}$ & 1.40 & $7.08^{+0.02}_{-0.02}$ & $44.45^{+0.02}_{-0.01}$ & $44.56^{+0.01}_{-0.02}$ \\
     F01076$-$1707 & 1.15 & 1.12 & $-0.04^{+0.09}_{-0.09}$ & 1.30 & $8.23^{+0.02}_{-0.02}$ & $44.15^{+0.10}_{-0.14}$ & $45.27^{+0.01}_{-0.01}$ \\
     F01173$+$1405 & 1.18 & 1.10 & $ 0.26^{+0.03}_{-0.03}$ & 1.40 & $7.94^{+0.04}_{-0.02}$ & $43.87^{+0.09}_{-0.07}$ & $45.24^{+0.01}_{-0.01}$ \\
     F01325$-$3623 & 1.40 & 1.11 & $ 0.25^{+0.03}_{-0.04}$ & 0.90 & $8.06^{+0.02}_{-0.02}$ & $41.24^{+0.87}_{-1.04}$ & $44.70^{+0.01}_{-0.01}$ \\
     F02208$+$4744 & 1.14 & 1.16 & $ 0.06^{+0.04}_{-0.05}$ & 1.08 & $7.88^{+0.02}_{-0.02}$ & $42.34^{+0.18}_{-0.27}$ & $44.73^{+0.01}_{-0.01}$ \\
     F02437$+$2122 & 1.10 & 1.00 & $ 0.31^{+0.04}_{-0.03}$ & 1.18 & $7.66^{+0.02}_{-0.02}$ & $43.09^{+0.06}_{-0.10}$ & $44.69^{+0.01}_{-0.01}$ \\
     F03117$+$4151 & 1.23 & 1.14 & $ 0.04^{+0.14}_{-0.21}$ & 1.00 & $8.11^{+0.06}_{-0.04}$ & $44.13^{+0.07}_{-0.07}$ & $44.93^{+0.02}_{-0.02}$ \\
     F04118$-$3207 & 1.26 & 1.11 & $-0.08^{+0.10}_{-0.13}$ & 1.08 & $7.71^{+0.02}_{-0.02}$ & $43.69^{+0.04}_{-0.06}$ & $44.59^{+0.01}_{-0.01}$ \\
\ph{F}04271$+$3849 & 1.25 & 1.09 & $ 0.13^{+0.06}_{-0.06}$ & 1.00 & $7.89^{+0.03}_{-0.02}$ & $42.11^{+0.61}_{-1.52}$ & $44.70^{+0.02}_{-0.01}$ \\
     F04454$-$4838 & 1.00 & 1.14 & $ 0.99^{+0.05}_{-0.05}$ & 1.40 & $8.01^{+0.02}_{-0.02}$ & $43.57^{+0.20}_{-0.21}$ & $45.39^{+0.01}_{-0.01}$ \\
\enddata
\tablecomments{
(1) Source name.
(2) The scaling factor to match the SL to LL spectra.
(3) The scaling factor to match the IRS spectra to \w4 band.
(4) The MIR extinction indicated as the optical depth at 9.7 \micron.
(5) The \umin\ of the interstellar radiation field.  The uncertainties
are not resolved for most of the object, so they are not listed.
(6) The dust mass.
(7) The integrated luminosity of the dust torus at 8--1000 \micron.
(8) The integrated luminosity of the host galaxy cold dust emission at
8--1000 \micron.
\\
(This table is available in its entirety in machine-readable form.)
}
\end{deluxetable*}

\end{document}